\documentclass[11pt]{article}
\usepackage[top=1in,left=1in,right=1in,bottom=1in]{geometry}
\pdfoutput=1

\newfont{\mycrnotice}{ptmr8t at 7pt}
\newfont{\myconfname}{ptmri8t at 7pt}
\let \originalleft \left
\let\originalright\right
\renewcommand{\left}{\mathopen{}\mathclose\bgroup\originalleft}
\renewcommand{\right}{\aftergroup\egroup\originalright}

\usepackage{color,array}
\usepackage{amsthm}
\usepackage{paralist}
\usepackage{booktabs}
\usepackage{paralist}

\usepackage[colorlinks=true, citecolor=black, urlcolor=cyan,
linkcolor=black]{hyperref}
\usepackage{amsmath,algorithmicx,amssymb,subfigure,graphicx,url,multirow}
\usepackage{enumerate}
\usepackage{algorithm}
\usepackage[noend]{algpseudocode}
\usepackage[bf,small,aboveskip=2pt,belowskip=0pt]{caption}

\DeclareCaptionType{copyrightbox}

\usepackage[compact]{titlesec}
\titlespacing*{\section}{0pt}{3pt}{3pt}
\titlespacing*{\subsection}{0pt}{2pt}{2pt}

\newcommand{\treeplus}{$C$-tree}
\newcommand{\treepluses}{$C$-trees}

\newcommand{\versionedgraph}{\mathsf{versioned\_graph}}
\newcommand{\version}{\mathsf{version}}
\newcommand{\versions}{\mathsf{versions}}
\newcommand{\vertex}{\mathsf{vertex}}

\newcommand{\buildgraph}{\textproc{BuildGraph}}
\newcommand{\insertedges}{\textproc{InsertEdges}}
\newcommand{\deleteedges}{\textproc{DeleteEdges}}
\newcommand{\insertvertices}{\textproc{InsertVertices}}
\newcommand{\deletevertices}{\textproc{DeleteVertices}}

\newcommand{\acquire}{\textproc{acquire}}
\newcommand{\verset}{\textproc{set}}
\newcommand{\release}{\textproc{release}}

\newcommand{\numvertices}{\textproc{NumVertices}}
\newcommand{\numedges}{\textproc{NumEdges}}
\newcommand{\findvertex}{\textproc{FindVertex}}

\newcommand{\vtxdegree}{\textproc{Degree}}
\newcommand{\vtxmap}{\textproc{Map}}
\newcommand{\vtxintersection}{\textproc{Intersection}}

\newcommand{\vset}{vertexSubset}

\newcommand{\emap}{\textproc{edgeMap}}

\newcommand{\codevar}[1]{\mathit{#1}}

\DeclareMathAlphabet{\mathbfsf}{\encodingdefault}{\sfdefault}{bx}{n}

\usepackage{amsmath,algorithmicx,amssymb,subfigure,graphicx,url,multirow}
\usepackage{algorithm}
\usepackage[noend]{algpseudocode}
\usepackage{enumitem}

\newcommand{\defn}[1]{\emph{\textbf{#1}}}

\newcommand{\myparagraph}[1]{\vspace{1.5pt}\noindent {\bf #1.}}

\newcommand{\id}[1]{\ifmmode\mathit{#1}\else\textit{#1}\fi}
\newcommand{\const}[1]{\ifmmode\mbox{\textc{#1}}\else\textsc{#1}\fi}
\newcommand{\plus}{prefix}
\newcommand{\Plus}{Prefix}
\newcommand{\pluses}{prefixes}

\newcommand{\plusnode}{prefix}

\newcommand{\tail}{tail}
\newcommand{\tails}{tails}
\newcommand{\chunk}{chunk}
\newcommand{\chunks}{chunks}

\newcommand{\heads}{heads}
\newcommand{\element}{element}
\newcommand{\elements}{elements}

\newcommand{\vtree}{vertex-tree}
\newcommand{\etree}{edge-tree}

\newcommand{\stinger}{Stinger}
\newcommand{\llama}{LLAMA}

\newtheorem{theorem}{Theorem}[section]
\newtheorem{corollary}{Corollary}[theorem]
\newtheorem{lemma}[theorem]{Lemma}

\usepackage{microtype}
\begin{document}
\date{}

\title{Low-Latency Graph Streaming Using Compressed Purely-Functional
  Trees\footnote{This is the full version of the paper appearing in the ACM
    SIGPLAN conference on Programming Language Design and Implementation (PLDI), 2019.}}

\author{
  Laxman Dhulipala\\ CMU\\ ldhulipa@cs.cmu.edu\and
  Guy E. Blelloch\\CMU\\guyb@cs.cmu.edu \and
  Julian Shun\\MIT CSAIL\\jshun@mit.edu}

\maketitle

\begin{abstract}

  Due to the dynamic nature of real-world graphs, there has been a growing
  interest in the graph-streaming setting where a continuous stream of
  graph updates is mixed with arbitrary graph queries.
  In principle, purely-functional trees are an ideal choice for this setting due as they enable safe
  parallelism, lightweight snapshots, and strict serializability for
  queries.  However, directly using them for graph processing would lead to significant space overhead and poor cache locality.

  This paper presents \treeplus{}s, a compressed purely-functional
  search tree data structure that significantly improves on the space usage
  and locality of purely-functional trees. The key idea is to use a
  chunking technique over trees in order to store multiple entries per
  tree-node. We design theoretically-efficient and practical
  algorithms for performing batch updates to \treepluses{}, and also
  show that we can store massive dynamic real-world graphs using only
  a few bytes per edge, thereby achieving space usage close to that of
  the best static graph processing frameworks.  

  To study the efficiency and applicability of our data structure, we
  designed Aspen, a graph-streaming framework that extends the 
  interface of Ligra with operations for updating graphs.  We show
  that Aspen is faster than two state-of-the-art
  graph-streaming systems, Stinger and LLAMA, while requiring less memory, and is
  competitive in performance with the state-of-the-art static graph
  frameworks, Galois, GAP, and Ligra+.  With Aspen, we are able to
  efficiently process the largest publicly-available graph with over
  two hundred billion edges in the graph-streaming setting using a single commodity multicore server
  with 1TB of memory.


\end{abstract}

\clearpage
\section{Introduction}\label{sec:intro}

In recent years, there has been growing interest in programming
frameworks for processing streaming graphs due to the fact that many
real-world graphs change in real-time (e.g.,~\cite{ediger2012stinger,
  feng2015distinger, green2016custinger, cheng2012kineograph,
  macko2015llama, Yin2018}). These graph-streaming systems receive a
stream of queries and a stream of updates (e.g., edge and vertex
insertions and deletions, as well as edge weight updates) and must
process both updates and queries with low latency, both in terms of
query processing time and the time it takes for updates to be
reflected in new queries.
There are several existing graph-streaming frameworks, such as
STINGER, based on maintaining a single mutable copy of the graph in
memory~\cite{ediger2012stinger, feng2015distinger,
  green2016custinger}.  Unfortunately, these frameworks require either
blocking queries or updates so that they are not concurrent, or giving
up serializability~\cite{Yin2018}.  Another approach is to use
snapshots~\cite{cheng2012kineograph, macko2015llama}.  Existing
snapshot-based systems, however, are either very space-inefficient, or
suffer from high latency on updates. Therefore, an important question
is whether we can design a data structure that supports lightweight
snapshots which can be used to concurrently process queries and
updates, while ensuring that the data structure is safe for
parallelism and achieves good asymptotic and empirical performance.


In principle, representing graphs using \emph{purely-functional
  balanced search trees}~\cite{abelson96sicp,okasaki98purely} can
satisfy both criteria.  Such a representation can use a search tree
over the vertices (the vertex-tree), and for each vertex store a
search tree of its incident edges (an edge-tree).  Because the trees
are purely-functional, acquiring an immutable snapshot is as simple as
acquiring a pointer to the root of the vertex-tree. Updates can then
happen concurrently without affecting the snapshot.  In fact, any
number of readers (queries) can concurrently acquire independent
snapshots without being affected by a writer.  A writer can make an
individual or bulk update and then set the root to make the changes
immediately and atomically visible to the next reader without
affecting current active readers.  A single update costs $O(\log n)$
work, and because the trees are purely-functional it is relatively
easy and safe to parallelize a bulk update.

However, there are several challenges that arise when comparing
purely-functional trees to compressed sparse row (CSR), the standard
data structure for representing static graphs in shared-memory graph
processing~\cite{saad2003iterative}. In CSR, the graph is stored as an array of vertices and an
array of edges, where each vertex points to the start of its edges in
the edge-array.  Therefore, in the CSR format, accessing all edges
incident to a vertex $v$ takes $O(deg(v))$ work, instead of $O(\log n + deg(v))$ work
for a graph represented using trees. Furthermore, the format requires only one pointer
(or index) per vertex and edge, instead of a whole tree node.
Additionally, as edges are stored contiguously, CSR has good cache
locality when accessing the edges incident to a vertex, while tree
nodes could be spread across memory. Finally, each set of edges can be
compressed internally using graph compression
techniques~\cite{shun2015ligraplus}, allowing massive graphs to be
stored using just a few bytes per
edge~\cite{dhulipala2018theoretically}.  This approach cannot be used
directly on trees.    This would all seem to put a search tree
representation at a severe disadvantage.

In this paper, we describe a compressed purely-functional tree data
structure that we call a \treeplus{}, which addresses the poor
space usage and locality of purely-functional trees. The \treeplus{}
data structure allows us to take advantage of graph compression
techniques, and thereby store very large graphs on a single machine.
The key idea of a \treeplus{} is to chunk the elements represented by
the tree and store each chunk contiguously in an array. Because
elements in a chunk are stored contiguously, the structure achieves
good locality. By ensuring that each chunk is large enough, we
significantly reduce the space used for tree nodes. Although the idea
of chunking is intuitive, designing a chunking scheme which admits
asymptotically-efficient algorithms for batch-updates and also
performs well in practice is challenging. We note that our chunking
scheme is independent of the underlying balancing scheme used, and
works for any type of element. In the context of graphs, because each
chunk in a \treeplus{} stores a sorted set of integers, we can
compress by applying
difference coding within each block and integer code the differences.
We compare to some other chunking schemes, including
$B$-trees~\cite{Bayer1972} and ropes~\cite{Acar14,Fluet08,Boehm95,yi08}
in Section~\ref{subsec:otherapproach}.

To address the asymptotic complexity issue, we observe that for many
graph algorithms the $O(\log n)$ work overhead to access vertices can
be handled in one of two ways.  The first is for global graph
algorithms, which process all vertices and edges. In this case, we can
afford to compute a \emph{flat snapshot}, which is an array of
pointers to the edge-tree for each vertex.  We show how to create a
flat snapshot using $O(n)$ work, $O(\log n)$ depth, and $O(n)$ space.
A flat snapshot can be created concurrently with updates and other
reads since it copies from the persistent functional representation.
Once a flat snapshot is created, the work for accessing the edges for a
vertex $v$ is only $O(deg(v))$, as with CSR. The second case is for
local graph algorithms, where we cannot afford to create a flat
snapshot. In this setting, we note that many local algorithms examine
all edges incident to a vertex after retrieving it.  Furthermore,
although real-world graphs are sparse, their average degree is often
in the same range or larger than $\log n$.  Therefore, the cost of
accessing a vertex in the vertex-tree can be amortized against the
cost of processing its incident edges.



To evaluate our ideas, we describe a new graph-streaming framework
called \emph{Aspen} that enables concurrent, low-latency processing of
queries and updates on graphs with billions of vertices and hundreds
of billions of edges, all on a relatively modest shared-memory machine
equipped with 1TB of RAM. Our system is fully serializable and
achieves high throughput and performance comparable to
state-of-the-art static graph processing systems. Aspen extends the
interface proposed by Ligra~\cite{shun2012ligra} with operations for
updating the graph.  As a result, all of the algorithms implemented
using Ligra, including graph traversal algorithms, local graph
algorithms~\cite{shun2016parallel}, algorithms using
bucketing~\cite{dhulipala2017julienne}, and
others~\cite{dhulipala2018theoretically}, can be run using Aspen with
minor modifications.  To make it easy to build upon or compare with our
work in the future, we have made Aspen
publicly-available at {\small \url{https://github.com/ldhulipala/aspen}}.

Compared to state-of-the-art graph-streaming frameworks,
Aspen provides significant improvements both in memory usage
 (8.5--11.4x more memory-efficient than
Stinger~\cite{ediger2012stinger} and 1.9--3.3x more memory-efficient
than LLAMA~\cite{macko2015llama}), and algorithm performance
(1.8--10.2x faster than Stinger and 2.8--15.1x faster than LLAMA). Aspen
is also comparable to the fastest static graph processing frameworks,
including GAP~\cite{BeamerAP15} (Aspen is 1.4x faster on average),
Ligra+~\cite{shun2015ligraplus} (Aspen is 1.4x slower on average), and
Galois~\cite{Nguyen2013} (Aspen is 12x faster on average).  Compared
to Ligra+, which is one of the fastest static compressed
graph representations, Aspen only requires between
1.8--2.3x more space.

Our experiments show that adding a continuous stream of edges while
running queries does not affect query performance by more than 3\%.
Furthermore, the latency is well under a millisecond, and the update
throughput ranges from 11K--78K updates per second when performing
one update at a time to \emph{105M--442M updates per second} when
performing batches of updates. We show that our update rates are
an order of magnitude faster than the update rates achievable by
Stinger, even when using very small batches.



The contributions of this paper are as follows:
\begin{enumerate}[label=(\arabic*),topsep=0pt,itemsep=0pt,parsep=0pt,leftmargin=15pt]
  \item A practical compressed purely-functional data structure for
    search trees, called the \treeplus{}, with operations
    that have strong theoretical bounds on work and depth.

  \item The approach of flat-snapshotting for \treepluses{} to reduce the cost
    of random access to the vertices of a graph.

  \item Aspen, a multicore graph-streaming framework built using
    \treeplus{}s that enables concurrent, low-latency processing of
    queries and updates, along with several algorithms using the
    framework.

  \item An experimental evaluation of Aspen in a variety of regimes
    over graph datasets at different scales, including the largest
    publicly-available graphs (graphs with billions of vertices and
    hundreds of billions of edges), showing significant improvements
    over state-of-the-art graph-streaming frameworks, and modest
    overhead over static graph processing frameworks.
\end{enumerate}

\section{Preliminaries}\label{sec:prelims}


\myparagraph{Notation and Primitives} We denote a graph by $G(V, E)$,
where $V$ is the set of vertices and $E$ is the set of edges in the
graph. For weighted graphs, the edges store real-valued weights.  The
number of vertices in a graph is $n = |V|$, and the number of edges is
$m = |E|$. Vertices are assumed to be indexed from $0$ to $n-1$. For
undirected graphs, we use $N(v)$ to denote the neighbors of vertex $v$
and $\emph{deg}(v)$ to denote its degree.
We assume that we have access to a family of uniformly (purely) random hash
functions which we can draw from in $O(1)$ work~\cite{CLRS,
pagh2008uniform}. In functions from such a family, each key is mapped to an
element in the range with equal probability, independent of the values that
other keys hash to, and the function can be evaluated for a given key in $O(1)$
work.

\myparagraph{Work-Depth Model}
We analyze algorithms  in the work-depth model, where the
\defn{work} is the number of operations used by the algorithm and the
\defn{depth} is the length of the longest sequential dependence in the
computation~\cite{JaJa92,BM10}.

\myparagraph{Purely-Functional Trees}
Purely-functional (mutation-free) data structures preserve previous
versions of themselves when modified and yield a new structure
reflecting the update~\cite{okasaki98purely}. The trees studied in
this paper are binary search trees, which represent a set of ordered
\elements{}. In a purely-functional tree, each \element{} is used as a
\emph{key}, and is stored in a separate tree node. The \elements{} can be
optionally associated with a value, which is stored in the node
along with the key. Trees can also be augmented with an associative
function $f$ (e.g., $+$), allowing the sum with respect to $f$ in a
range of the tree be queried in $O(\log n)$ work and depth, where $n$ is the
number of \elements{} in the tree.

\myparagraph{Interfaces for Graphs}
We will extend the interface defined by Ligra~\cite{shun2012ligra} and so we
review its interface here.
We use the \defn{\vset{}} data structure which represents
subsets of vertices, and the \emap{} primitive which is used for mapping over
edges incident to sets of vertices. \emap{} takes as input a subset of vertices
and applies a function over the edges incident to the subset that satisfy a
condition (e.g., edges to vertices that have not yet been visited by a
breadth-first search). More precisely, \textsc{\textbf{edgeMap}} takes as input a graph
$G(V,E)$, a \vset{} $U$, and two boolean functions $F$ and $C$; it applies $F$
to $(u,v) \in E$ such that $u \in U$ and $C(v) = \codevar{true}$ (call this
subset of edges $E_{a}$), and returns a \vset{} $U'$ where $u \in U'$ if and
only if $(u,v) \in E_{a}$ and $F(u,v) = \codevar{true}$.

\section{Compressed Purely-Functional Trees}\label{sec:trees}

In this section, we describe a compressed purely-functional search tree data
structure which we refer to as a \defn{\treeplus{}}. After describing the
data structure in Section~\ref{sec:compressedtrees}, we argue that our
design improves locality and reduces space-usage relative to ordinary
purely-functional trees (Section~\ref{subsec:compression}). Finally,
we compare the \treeplus{} data structure to other possible design
choices, such as $B$-trees (Section~\ref{subsec:otherapproach}).



\begin{figure}[t]
  \centering
    \includegraphics[width=0.5\textwidth]{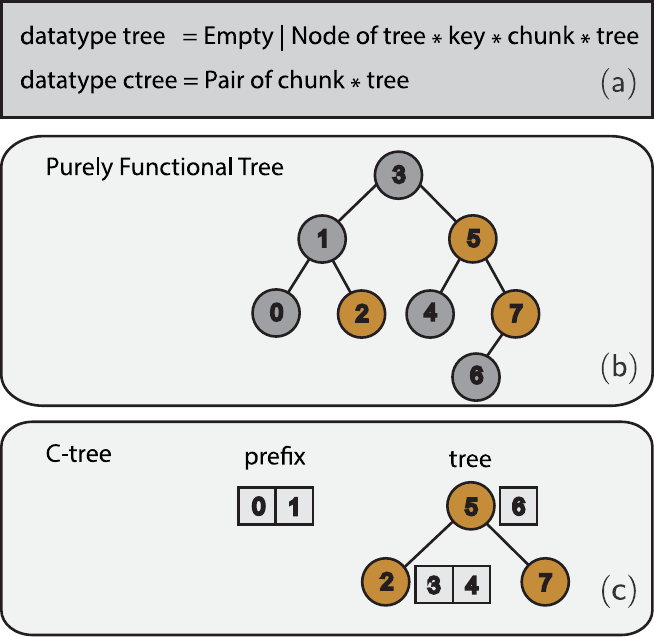}\\ 
      \vspace{0.5em}
    \caption{
      This figure gives the definition of the \treeplus{} data
      structure in an ML-like language (subfigure (a)) and illustrates
      the difference between a purely-functional tree and a
      \treeplus{} when representing a set of integers, $S$. Subfigure
      (b) shows a purely-functional tree where each \element{}
      in $S$ is stored in a separate tree node. We color the \elements{} in $S$
      that are sampled as heads yellow, and color the
      non-head \elements{} gray.  Subfigure (c) illustrates
      how the \treeplus{} stores $S$, given the heads. Notice that the
      \treeplus{} has a chunk (the \plus{}) which contains non-head
      \elements{} that are not associated with any head, and that each
      head stores a chunk (its \tail{}) containing all non-head
      \elements{} that follow it until the next head.
  }\label{fig:c-tree-defn}
\end{figure}

\subsection{\treeplus{} Definition}\label{sec:compressedtrees}
The main idea of \treepluses{} is to apply a chunking scheme over the
tree to store multiple \elements{} per tree-node.  The chunking scheme
takes the ordered set of \elements{} to be represented and ``promotes''
certain \elements{} to be heads, which are stored in a tree.  The
remaining \elements{} are stored in tails associated with each tree node.  To
ensure that the same keys are promoted in different trees, a hash function
is used to choose which \elements{} are promoted.
An important goal for \treepluses{} is
to maintain similar asymptotic cost bounds as for the uncompressed
trees while improving space and cache performance, and to this end we
describe theoretically efficient implementations of tree primitives in
Section~\ref{sec:ctreeops}.

\myparagraph{More formally} For an \element{} type $K$, fix a hash
function, $h : K \rightarrow \{1, \ldots N\}$, drawn from a uniformly random
family of hash functions ($N$ is some sufficiently large
range).
Let $b$ be a \emph{chunking parameter}, a constant which
controls the granularity of the chunks.
Given a set $E$ of $n$ \elements{}, we first compute the set of
\defn{heads} $H(E) = \{e \in E\ |\ h(e) \mod b = 0\}$.  For each $e
\in H(E)$ let its \defn{tail} be $t(e) = \{x \in E~|~ e < x <
next(H(E),e)\}$, where $next(H(e),e)$ returns the next
\element{} in $H(E)$ greater than $e$.  We then construct a
purely-functional tree with keys $e \in H(E)$ and associated values
$t(e)$.

Thus far, we have described the construction of a tree over the head
\elements{}, and their tails.  However, there may be a ``tail'' at the
beginning of $E$ that has no associated head, and is therefore not
part of the tree. We refer to this chunk of \elements{} as the
\defn{\plus}.  We refer to either a \tail{} or \plus{} as a
\defn{\chunk{}}.  We represent each \chunk{} as a (variable-length)
array of \elements{}.  As described later, when the \elements{} are integers
we can use difference encoding to compress each of the chunks.  The
overall \defn{\treeplus{}} data structure consists of the tree over
head keys and tail values, and a single (possibly empty) \plus.
Figure~\ref{fig:c-tree-defn} illustrates the \treeplus{} data
structure over a set of integer \elements{}.

\myparagraph{Properties of \treepluses{}}
The expected size of chunks in a \treeplus{} is $b$ as
each \element{} is independently selected as a head under $h$ with
probability $1/b$. Furthermore, the
chunks are unlikely to be much larger than $b$---in particular, a
simple calculation shows that the chunks have size at most $O(b \log
n)$ with high probability (w.h.p.),\footnote{ We use \defn{with high
    probability (w.h.p.)} to mean with probability $1 - 1/n^c$ for
  some constant $c>0$.}  where $n$ is the number of \elements{} in the
tree. Notice that an \element{} chosen to be a head will be a head in any
\treepluses{} containing it, a property that simplifies the
implementation of primitives on \treepluses{}.

Our chunking scheme has the following bounds, which we prove in
Appendix~\ref{subsec:properties}.


\vspace{-3pt}

\begin{lemma}\label{lem:structure}
The number of heads (keys) in a \treeplus{} over a set $E$ of $n$
\elements{} is $O(n/b)$ w.h.p. Furthermore,
the maximum size of a tail (the non-head nodes associated with a head)
or \plus{} is $O(b \log n)$ w.h.p.
\end{lemma}


\vspace{-3pt}
We also obtain the following corollary.
\vspace{-3pt}
\begin{corollary}\label{cor:height}
  When using a balanced binary tree for the heads (one with $O(
  \log n)$ height for $n$ keys), the height of a \treeplus{} over a
  sequence $E$ of $n$ \elements{} is $O(\log (n/b))$ w.h.p.
\end{corollary}

\subsection{\treeplus{} Compression}\label{subsec:compression}
In this section, we first discuss the improved space usage of
\treepluses{} relative to purely-functional trees without any
assumption on the underlying type of \elements{}. We then discuss how we
can further reduce the space usage of the data structure in the case
where the \elements{} are integers.


\myparagraph{Space Usage and Locality}
Consider the layout of a \treeplus{} compared to a purely-functional
tree.
By Lemma~\ref{lem:structure}, the
expected number of heads is $O(n/b)$.
Therefore, compared to a purely-functional tree, which allocates $n$
tree nodes, we reduce the number of tree nodes allocated by a factor
of $b$. As each tree node is quite large (in our implementation, each
tree node is at least 32 bytes), reducing the number of nodes by a
factor of $b$ can significantly reduce the size of the tree.
Experimental results are given in Section~\ref{sec:aspcompression}.




In a
purely-functional tree, in the worst case,
accessing each \element{} will incur a cache miss, even in the case where
\elements{} are smaller than the size of a cache line.
In a \treeplus{}, however, by choosing $b$, the chunking parameter, to
be slightly larger than the cache line size ($\approx 128$),
we can store
multiple \elements{} contiguously within a single \chunk{} and amortize
the cost of a cache miss across all \elements{} read from the \chunk{}.
Furthermore, note that the data structure can provide locality
benefits even in the case when the size of an \element{} is larger than
the cache line size, as a modest value of $b$ will ensure that reading
all but the heads, which constitute an $O(1/b)$ fraction of the
\elements{}, will be contiguous loads from the \chunks{}.

\myparagraph{Integer \treepluses{}} In the case where the \elements{} are
integers, the \treeplus{} data structure can exploit the fact that
\elements{} are stored in sorted order in the \chunks{}  to
further compress the data structure.
We apply a \emph{difference encoding} scheme to each
\chunk{}. Given a \chunk{} containing $d$ integers, $\{I_{1}, \ldots,
I_{d}\}$, we compute the differences $\{I_{1}, I_{2} - I_{1},
\ldots, I_{d} - I_{d-1}\}$. The differences are then encoded using a
byte-code~\cite{shun2015ligraplus,Witten1999}.
We applied
byte-codes due to the fact that they are fast to decode while
achieving most of the memory savings that are possible using a shorter
code~\cite{blandford2004experimental,Witten1999}.

Note that in the common case when $b$ is a constant, the size of each
chunk is small ($O(\log n)$ w.h.p.). Therefore, despite the fact that
each \chunk{} must be processed sequentially, the cost of the
sequential decoding does not affect the overall work or depth of
parallel tree methods. For example, mapping over all \elements{} in the
\treeplus{}, or finding a particular \element{} have the same asymptotic
work as purely-functional trees and optimal ($O(\log n)$) depth. To make the
data structure dynamic, \chunks{} must also be recompressed when updating a
\treeplus{}, which has a similar cost to decompressing the \chunks{}. In the
context of graph processing, the fact that methods over a \treeplus{} are easily
parallelizable and have low depth lets us avoid designing and implementing a
more complex parallel decoding scheme, like the parallel byte-code in
Ligra+~\cite{shun2015ligraplus}.


\begin{figure}[!t]
  \centering
  \includegraphics[width=0.55\textwidth]{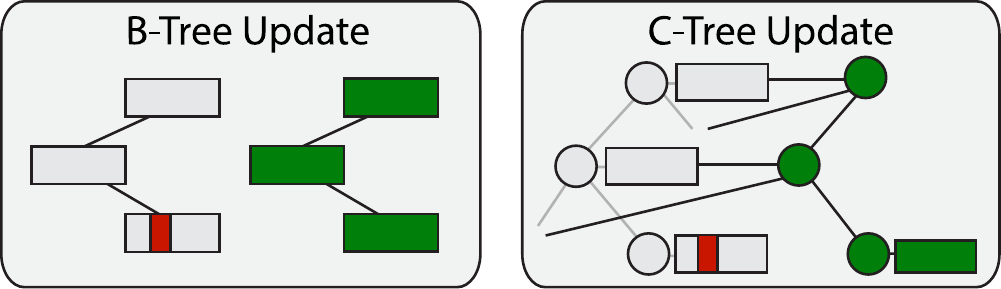}\\ 
  \vspace{0.5em}
    \caption{
      This figure shows the difference between performing a single
      update in a $B$-Tree versus an update in a \treeplus{}. The data
      marked in green is newly allocated in the update. Observe that
      updating single element in a \treeplus{} in the worst-case
      requires copying a path of nodes, and copying a single \chunk{}
      if the element is not a head. Updating an element in a $B$-tree
      requires copying $B$ pointers (potentially thousands of bytes)
      per level of the tree, which adds significant overhead in terms
      of memory and running time.
  }\label{fig:ctree-update}
\end{figure}

\subsection{Other Approaches}\label{subsec:otherapproach}
Our data structure is loosely based on a previous sequential approach to
chunking~\cite{BB04}. That approach was designed to be a generic addition to any
existing balanced tree scheme for a dictionary and has
overheads due to this goal.

Another option is to use $B$-trees~\cite{Bayer1972}.  However, the
objective of a $B$-tree is to reduce the height of a search tree to
accelerate searching a tree in external memory, whereas our goal is to
build a data structure that stores many contiguous segments in a
single node to make compression possible.  The problem with $B$-trees
in our purely-functional setting is that we require path copying
during functional updates, as illustrated in Figure~\ref{fig:ctree-update}.  In our trees, this only requires copying a
single binary node (32 or 40 bytes in our implementation) per level of
the tree.  For a $B$-tree, it would require copying $B$ pointers
(potentially thousands of bytes) per level of the tree, adding
significant overhead in terms of memory and running time.

There is also work on chunking of functional trees for
representing strings or (unordered)
sequences~\cite{Acar14,Fluet08,Boehm95,yi08}.  The motivation is
similar (decrease space and increase locality), but the fact they are
sequences rather than search trees makes the tradeoffs different.
None of this work uses the idea of hashing or efficiently searching the trees.
Using a hash function to select the heads has an important advantage
in simplifying much of the code, and proving asymptotic bounds.
Keeping the \elements{} with internal nodes and using a prefix allows
us to access the first $b$ \elements{} (or so) in constant work.


\section{Operations on \treepluses{}}\label{sec:ctreeops}

In this section, we show how to support various tree operations over
\treepluses{}, such as building, searching and performing
batch-updates to the data structure.  These are operations that we
will need for efficiently processing and updating graphs.  We argue
that the primitives are theoretically efficient by showing bounds on
the work and depth of each operation. We also describe how to support
augmentation in the data structure using an underlying augmented
purely-functional tree. We note that the \treeplus{} interfaces
defined in this section operate over \element{}-value pairs, whereas the
\treepluses{} defined in Section~\ref{sec:compressedtrees} only stored
a set of \elements{} for the sake of illustration. The algorithm
descriptions elide the values associated with each \element{} for the
sake of clarity. We use operations on an underlying purely-functional
tree data structure in our description, and state the bounds for
operations on these trees as necessary (e.g., the trees described in
Blelloch et al.~\cite{blelloch16justjoin} and Sun et
al.~\cite{sun2018pam}).  The primitives in this section for a
\treeplus{} containing \elements{} of type $E$ and values of type $V$ are
defined as follows.

\begin{itemize}[topsep=0pt,itemsep=0pt,parsep=0pt,leftmargin=8pt]
  \item \defn{Build$(S, f_{V})$} takes a sequence of \element{}-value
    pairs and returns a \treeplus{} containing the \elements{} in $S$
    with duplicate values combined using a function $f_{V} : V \times
    V \rightarrow V$.

  \item \defn{Find$(T, e)$} takes a \treeplus{} $T$ and an \element{} $e$
    and returns the entry of the largest \element{} $e' \leq e$.

  \item \defn{Map$(T, f)$} takes a \treeplus{} $T$ and a function $f :
    V \rightarrow ()$ and applies $f$ to each \element{} in $T$.

  \item \defn{MultiInsert}($T, f, S$) and \defn{MultiDelete}($T, S$)
    take a \treeplus{} $T$, (possibly) a function $f : V \times V \rightarrow V$
    that specifies how to combine values, and a sequence $S$ of
    \element{}-value pairs, and returns a \treeplus{} containing the
    union or difference of $T$ and $S$.


\end{itemize}

\noindent Our algorithms for \textsc{Build}, \textsc{Find}, and
\textsc{Map} are straightforward, so due to space constraints, we give
details about these implementations in Appendix~\ref{subsec:primdetails}.

\begin{figure*}[!t]
  \centering
    \vspace{0.5em}
    \includegraphics[width=\textwidth]{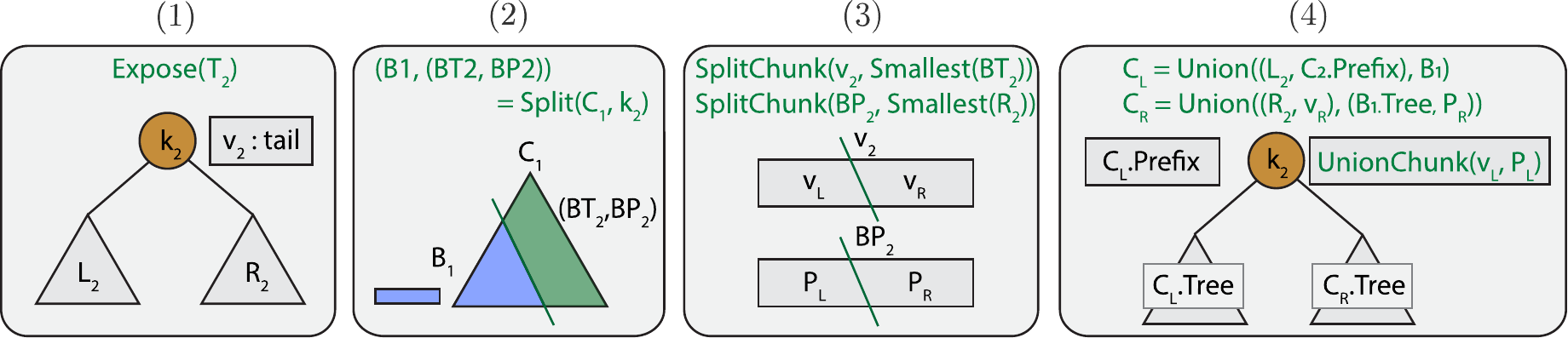}\\ 
    \vspace{0.5em}
    \caption{
      \small This figure illustrates how our \textsc{Union} algorithm
      computes the union of two \treepluses{}, $T_1$ and $T_2$. The
      text at the top of each figure (in green) denotes the
      sub-routine that is called, and the bottom portion of the figure
      illustrates the output of the call.}\label{fig:c-tree-union}
\end{figure*}

\subsection{Algorithms for Batch Insertions and Deletions}\label{subsec:union}
Our \textsc{MultiInsert} and \textsc{MultiDelete} algorithms are based
on more fundamental algorithms for \textsc{Union},
\textsc{Intersection}, and \textsc{Difference} on \treepluses{}. Since
we can simply build a tree over the input sequence to
\textsc{MultiInsert} and call \textsc{Union} (or \textsc{Difference}
for \textsc{MultiDelete}), we focus only on the set operations.
Furthermore, because the algorithms for \textsc{Intersection} and
\textsc{Difference} are conceptually very similar to the algorithm for
\textsc{Union}, we only describe in detail the \textsc{Union} algorithm, and
\textsc{Split}, an important primitive used to implement \textsc{Union}.

\begin{algorithm}[!t]
\caption{\textsc{Union}} \label{alg:union-rec}
\small
\begin{algorithmic}[1]
\Function{Union}{$C_{1}, C_{2}$}
\State \hspace{-1em} {\bf case} $(C_1, C_2)$\ {\bf of}
\State $((\mathsf{null}, \_), \_) \rightarrow \textsc{UnionBC}(C_{1}, C_{2})$ \label{line:bc1}
\State $(\_, (\mathsf{null}, \_)) \rightarrow \textsc{UnionBC}(C_{2}, C_{1})$ \label{line:bc2}
\State $((T_1, P_1), (T_2, P_2)) \rightarrow$
\State \hspace{-0.8em}{\bf let}
  \State $\mathsf{val}$\ $(L_2, k_2, v_2, R_2) = \textsc{Expose}(T_2)$\label{line:expose}
  \State $\mathsf{val}$\ $(B_1, (BT_{2}, BP_{2})) = \textsc{Split}(C_{1}, k_2)$\label{line:split}
  \State $\mathsf{val}$\ $(v_{L}, v_{R}) = \textsc{SplitChunk}(v_2,
  \textsc{Smallest}(BT_{2}))$ \label{line:splitv2}
  \State $\mathsf{val}$\ $(P_{L}, P_{R}) = \textsc{SplitChunk}(BP_{2},
  \textsc{Smallest}(R_{2}))$ \label{line:splitb2prefix}
\State $\mathsf{val}$\ $v_2' = \textsc{UnionChunk}(v_{L}, P_{L})$ \label{line:unionv2list}
\State $\mathsf{val}$\ $(C_{L}, C_{R}) = \textsc{Union}(B_{1}, (L_{2}, P_{2}))\ ||\ $
\Statex $\hspace{7.55em}\textsc{Union}((BT_{2}, P_{R}), (R_{2}, v_{R}))$ \label{line:unionrec}
\State \hspace{-0.8em}{\bf in}
\State $\mathsf{ctree}$$(\textsc{Join}(C_{L}.\textsc{Tree}, C_{R}.\textsc{Tree}, k_2, v_2'), C_{L}.\textsc{\Plus})$ \label{line:unionret}
\State \hspace{-0.8em}{\bf end}
\EndFunction
\end{algorithmic}
\end{algorithm}

\myparagraph{Union} Our \textsc{Union} algorithm
(Algorithm~\ref{alg:union-rec}) is based on the recursive algorithm
for \textsc{Union} given by Blelloch et
al.~\cite{blelloch16justjoin}. The main differences between the
implementations are how to split a \treeplus{} by a given \element{},
and how to handle \elements{} in the \tails{} and \pluses{}. The
algorithm takes as input two \treepluses{}, $C_1$ and $C_2$, and
returns a \treeplus{} $C$ containing the elements in the union of
$C_1$ and $C_2$. Figure~\ref{fig:c-tree-union} provides an
illustration of how our \textsc{Union} algorithm computes the union of
two \treepluses{}. The algorithms use the following operations defined
on \treepluses{} and \chunks{}. The \textsc{\textbf{Expose}} operation
takes as input a tree and returns the left subtree, the \element{} and
\plus{} at the root of the tree, and the right subtree. The
\textsc{\textbf{Split}} operation takes as input a \treeplus{} $B$ and
an \element{} $k$, and returns two \treepluses{} $B_1$ and $B_2$,
where $B_1$ (resp.  $B_2$) are a \treeplus{} containing all
\elements{} less than (resp. greater than) $k$. It can also optionally return a
boolean indicating whether $k$ was found in $B$, which is used when implementing
\textsc{Difference} and \textsc{Intersection}. The \textsc{\textbf{Smallest}} operation returns the smallest head in a tree. The
\textsc{\textbf{UnionBC}} algorithm merges a \treeplus{} consisting
of a \plus{} and empty tree, and another \treeplus{}.  We also use the
\textsc{\textbf{SplitChunk}} and \textsc{\textbf{UnionChunk}} operations,
which are defined similarly to \textsc{Split} and \textsc{Union} for
\chunks{}.

The idea of the algorithm is to call \textsc{Expose} on the tree of one of the two
\treepluses{} ($C_2$), and split the other \treeplus{} ($C_1$) based
on the \element{} exposed at the root of $C_2$'s tree
(Line~\ref{line:expose}). The split on $C_1$ returns the trees $B_1$
and $B_2$ (Line~\ref{line:split}). The algorithm then recursively calls \textsc{Union}
on the \treepluses{} constructed from $L_2$ and $R_2$, the left and right
subtrees exposed in $C_2$'s tree with the \treepluses{} returned by
\textsc{Split}, $B_1$, and $B_2$.


However, some care must be taken, since \elements{} in $k_2$'s \tail{},
$v_2$, may come after some \heads{} in $B_2$. Similarly, \elements{} in
$B_2$'s prefix may come after some \heads{} of $R_2$. In both cases, we
should merge these \elements{} with their corresponding \heads{}' tails.
We handle these cases by splitting $v_2$ by the leftmost \element{} of
$B_2$ (producing $v_L$ and $v_R$), and splitting $B_2$'s prefix by the
leftmost \element{} of $R_2$ (producing $P_L$ and $P_R$). The left
recursive call to \textsc{Union} just takes the \treepluses{} $B_1$
and $(L_2, P_2)$.
The right recursive call takes the \treepluses{}
$(B_2.\textsc{Tree}, P_R)$, and $(R_2, v_R)$. Note that all \elements{}
in the \pluses{} $P_R$ and $v_R$ are larger than the smallest head in
$B_2$ and $R_2$. Therefore, the \treeplus{} returned from the right
recursive call has an empty \plus{}. The output of \textsc{Union} is
the \treeplus{} formed by joining the left and right trees from the
recursive calls, $k_2$, and the \tail{} $v_2'$ formed by unioning
$v_L$ and $P_L$, with the prefix from $C_L$.


\myparagraph{UnionBC}
Recall that the \textsc{UnionBC} algorithm merges a \treeplus{} consisting of a
\plus{} and empty tree, and another \treeplus{}.
We give a detailed description and pseudocode of the algorithm in
Appendix~\ref{subsec:primdetails}. The idea of \textsc{UnionBC} is
to split the \plus{} based on the leftmost \element{} of $P$'s tree into two
pieces, $P_{L}$ and $P_{R}$ containing \elements{} less than and greater than
the leftmost \element{} respectively. $P_{L}$ is merged with $P$'s
\plus{} to generate $P'$. The \elements{} in $P_{R}$ find the heads they
correspond to by searching the tree for the largest head that is
smaller than them. We then construct a sequence of head-\tail{} pairs
by inserting each \element{} in $P_{R}$ into its corresponding \elements{}
\tail{}. Finally, we generate a new tree, $T'$, by performing a
\textsc{MultiInsert} into $C$'s tree with the updated head-\tail{}
pairs. The return value is the \treeplus{} $(T', P')$.

\myparagraph{Split}
$\textsc{Split}$ takes a \treeplus{}, $C = (T, P)$, and an \element{} $k$
and returns a triple consisting of a \treeplus{} of all \elements{} less
than $k$, whether the \element{} was found, and a \treeplus{} of all
\elements{} greater than $k$. We provide a high-level description of the
algorithm here and defer the pseudocode and details to
Appendix~\ref{subsec:primdetails}.

The algorithm works by enumerating cases for how the split key can
split $C$. If $k$ is less than the first \element{} in $P$, then we
return an empty \treeplus{}, false, indicating that $k$ was not found,
and $C$ as the right \treeplus{}. Similarly, if $k$ splits $P$ (it
lies between the first and last \elements{} of $P$) then we split
$P$, and return the list of \elements{} less than the split key as the
left \treeplus{}, with the boolean and right tree handled similarly.
Otherwise, if the above cases did not match, and the tree is null,
then we return $C$ as the left \treeplus{}. The recursive cases are
similar to how \textsc{Split} is implemented in Blelloch et
al.~\cite{blelloch16justjoin}, except for the case where $k$ splits
the \tail{} at the root of the tree. Another important detail is how
we compute the first and last \elements{} of a \chunk{}. Instead of
scanning the \chunk{}, which will cause us to do work proportional to
the sum of \chunks{} on a root-to-leaf path in the tree, we store the
first and last \elements{} at the head of each \chunk{} to perform this
operation in $O(1)$ work and depth. This modification is important to
show that \textsc{Split} can be done in $O(b \log n)$ work and depth
w.h.p. on a \treeplus{}.

\subsection{Work and Depth Bounds}
Due to space constraints, we provide the details, correctness proofs,
and analysis for our \treeplus{} primitives in Appendix~\ref{subsec:primdetails}, and state the work and depth bounds below.





\myparagraph{Building}
Building (Build$(S, f_{V})$) a \treeplus{} can be done in $O(n\log n)$
work and $O(b \log n)$ depth w.h.p. for a sequence of length $n$.

\myparagraph{Searching} Searching (\textsc{Find}$(T, e)$) for an
\element{} $e$ in a \treeplus{} can be implemented in $O(b\log n)$
work and depth w.h.p., and $O(b + \log n)$ work and depth in expectation.

\myparagraph{Mapping}
Mapping (\textsc{Map}$(T, f)$) over a \treeplus{} containing $n$
\elements{} with a constant-work function $f$ can be done in $O(n)$ work
and $O(b \log n)$ depth w.h.p.

\myparagraph{Batch Updates}
Batch updates (\textsc{MultiInsert}($T, f, S$) and
\textsc{MultiDelete}($T, S$)) can be performed in $O(b^2(k \log ((n/k) +
1)))$ expected work and $O(b \log k \log n)$ depth w.h.p. where
$k = \min(|T|, |S|)$ and $n = \max(|T|, |S|)$.

\begin{figure*}[!t]
  \centering
    \includegraphics[width=\textwidth]{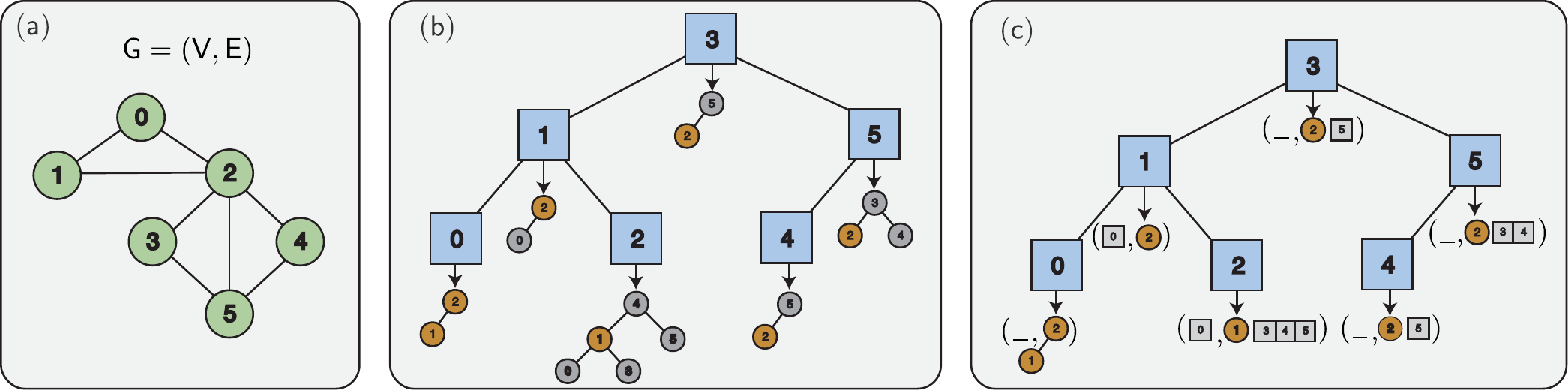}\\ 
    \vspace{0.5em}
    \caption{We illustrate how the graph (shown in subfigure
      (a)) is represented as a simple tree of trees (subfigure (b))
    and as a tree of \treepluses{} (subfigure (c)). As in
    Figure~\ref{fig:c-tree-defn}, we color \elements{} (in this case
    vertex IDs) that are sampled as heads  yellow. The \plus{}
    and tree in each \treeplus{} are drawn as a tuple, following the
    datatype definition in Figure~\ref{fig:c-tree-defn}.
  }\label{fig:tree-of-trees}
  \vspace{0.5em}
\end{figure*}

\section{Representing Graphs as Trees}\label{sec:graphsastrees}

\myparagraph{Representation}
An undirected graph can be implemented using purely functional
tree-based data structures by representing the set of vertices as a
tree, which we call the \defn{\vtree{}}. Each vertex in the \vtree{}
represents its adjacency information by storing a tree of identifiers
of its adjacent neighbors, which we call the \defn{\etree{}}. Directed
graphs can be represented in the same way by simply storing two
\etree{}s per vertex, one for the out-neighbors, and one for the
in-neighbors. The resulting graph data structure is a tree-of-trees
that has $O(\log n)$ overall depth using any balanced tree
implementation (w.h.p. using a treap). Figure~\ref{fig:tree-of-trees}
illustrates the \vtree{} and the \etree{}s for an example graph
(subfigure (a)). Subfigure (b) illustrates how the graph is
represented using simple trees for both the \vtree{} and \etree{}.
Subfigure (c) illustrates using a simple tree for the \vtree{} and a
\treeplus{} for the \etree{}. We augment the \vtree{} to store the
number of edges contained in its subtrees, which is needed to compute
the total number of edges in the graph in $O(1)$ work. Weighted graphs
can be represented using the same structure, with the only difference
being that the elements in the \etree{}s are modified to store an associated
edge weight. Note that computing associative functions over the
weights (e.g., aggregating the sum of all edge-weights) could be easily done by
augmenting the edge and vertex-trees.
We also note that the \vtree{} could also be compressed using
a \treeplus{} but defer evaluating this idea for future
work.


\myparagraph{Basic Graph Operations}
We can compute the number of vertices and number of edges in the graph
by querying the size (number of keys) in the \vtree{} and the
augmented value of the \vtree{} respectively, which can both be done
in $O(1)$ work. Finding a particular vertex just searches the
\vtree{}, which takes $O(\log n)$ work and depth.

\myparagraph{\emap{}} We implement \emap{} (defined in
Section~\ref{sec:prelims}) by mapping over the vertices in the input
\vset{} $U$ in parallel and for each vertex $u \in U$ searching the
\vtree{} for its edge-tree, and then mapping over $u$'s incident
neighbors, again in parallel.  For each of $u$'s neighbors $v$, we
apply the map function $F(u,v)$ if the filter function $C(v)$ returns
\emph{true}.  Other than finding vertices in the input \vset{} in $G$
and traversing edges via the tree instead of an array, the
implementation is effectively the same as in
Ligra~\cite{shun2012ligra}.  The direction
optimization~\cite{beamer12direction,shun2012ligra} can also be implemented, and we describe more details later in this
section.  Assuming the functions $F$ and $C$ take constant work,
\textsc{EdgeMap} takes $O(\sum_{u\in U}\emph{deg}(u) + |U|\log n)$
work and $O(\log n)$ depth.

\myparagraph{Batch Updates}
Inserting and deleting edges are defined similarly, and so we only
provide details for \textsc{InsertEdges}. Note that updates (e.g., to
the weight) of existing edges can be done within this interface. Let
$A$ be the sequence (batch) of edge updates and let $k = |A|$.

We first sort the batch of edge pairs using a comparison sort.
Next,
we compute an array of source vertex IDs that are being updated and
for each ID, in parallel, build a tree over its updated edges. We can
combine duplicate updates in the batch by using the
duplicate-combining function provided by the \treeplus{} constructor. As the
sequence is sorted, the build costs $O(k)$ work and $O(\log k)$ depth.
Next, in the update step, we call \textsc{MultiInsert} over the
\vtree{} with each $(source, tree)$ pair in the previous sequence. The
combine function for \textsc{MultiInsert} combines existing values
(\etree{}s) with the new \etree{}s by calling \textsc{Union} on the
old \etree{} and new \etree{}.

We give a simple worst-case analysis of the algorithm and show that
the algorithm performs $O(k \log n)$ work overall, and has $O(\log^3
n)$ depth. All steps before the \textsc{MultiInsert} cost $O(k\log k)$
work, and $O(\log k)$ depth in total, as they sort and apply parallel
sequence operations to sequences of length $k$~\cite{JaJa92}. As the depth of both
the \vtree{} and \etree{} is $O(\log n)$, the overall work of updating
both the \vtree{} and each affected \etree{} can be upper bounded by
$O(k \log n)$. The depth of \textsc{MultiInsert} is $O(\log n(\log m +
D_{\textsc{Union}}))$, where $D_{\textsc{Union}})$ is the depth of
union. This simplifies to $O(\log^3 n)$ by upper-bounding
$D_{\textsc{Union}}$ on any two trees as $O(\log^2 n)$, as shown in
Appendix~\ref{subsec:primdetails}.

\subsection{Efficiently Implementing Graph Algorithms}\label{sec:graphalgs}
We now address how to efficiently implement graph algorithms using
a tree of \treepluses{}, mitigating the increase in access times due
to using trees. We first describe a technique for handling the
asymptotic increase in work for global graph algorithms due to using
trees. We then consider local algorithms, and argue that for many
local algorithms, the extra cost of searching the \vtree{} can be
amortized.  Finally, we describe how direction
optimization~\cite{beamer12direction} can be easily implemented over
the \treeplus{} data structure.

\myparagraph{Flat Snapshots}
Notice that algorithms in our framework that use \emap{} incur an
extra $O(K \log n)$ factor in their work, where $K$ is the total
number of vertices accessed by \emap{} over the course of the
algorithm. For an algorithm like breadth-first search, which runs in
$O(m + n)$ work and $O(D\log n)$ depth for a graph with diameter $D$ using a
static-graph processing framework~\cite{dhulipala2018theoretically}, a
naive implementation using our framework will require performing $O(m
+ n\log n)$ work (the depth is the same, assuming that $b$ is a
constant).

Instead, for \emph{global graph algorithms}, which we loosely define
as performing $\Omega(n)$ work, we can afford to take a \emph{flat
snapshot} of the graph, which reduces the $O(K \log n)$ term to
$O(K)$. The idea of a flat snapshot is very simple---instead of
accessing vertices through the vertex-tree, and calling \textsc{Find}
for each $v$ supplied to \emap{}, we just precompute the pointers to the edge-trees
for all $v \in V$ and store them in an array of size
$n$.
This can be done in linear work and $O(\log n)$ depth by traversing the
vertex-tree once to fetch the pointers.
By providing this array, which we call a \defn{flat snapshot} to
each call to \emap{}, we can directly access the edges tree in $O(1)$
work and reduce the work of \emap{} on a \vset{}, $U$, to
$O(\sum_{u\in U}\emph{deg}(u) + |U|)$. In practice, using a flat snapshot speeds
up BFS queries on our input graphs by an average of 1.26x (see
Table~\ref{table:flat-snapshot-comparison}).


\myparagraph{Local Algorithms}
In the case of local graph algorithms, we often cannot afford to
create a flat snapshot without a significant increase in the work. We
observe, however, that after retrieving a vertex many local
algorithms will process all edges incident to it.  Because the average
degree in real-world graphs is often in the same range or
larger than $\log n$ (see Table~\ref{table:sizes}), the logarithmic
overhead of accessing a vertex in the vertex-tree in these graphs can
be amortized against the cost of processing the edges incident to the
vertex, on average.

\myparagraph{Direction Optimization} Direction optimization is a
technique first described for breadth-first search in Beamer et
al.~\cite{beamer12direction}, and later generalized as part of Ligra
in its \emap{} implementation~\cite{shun2012ligra}.  It combines a
sparse traversal, which applies the $F$ function in \emap{} to the
outgoing neighbors of the input \vset{} $U$, with a dense traversal,
which applies $F$ to the incoming neighbors $u$ of all vertices $v$ in
the graph where $C(v)=\emph{true}$ and $u \in U$. The dense traversal
improves locality for large input \vset{}s, and reduces edge
traversals in some algorithms, such as breadth-first search.  The
traversal mode on each iteration is selected based on the size of $U$
and its out-degrees. We implemented the optimization by implementing a sparse traversal and a dense
traversal that traverses the underlying \treepluses{}.

\section{Aspen Graph-Streaming Framework}\label{sec:aspen}

In this section, we outline the Aspen interface and implementation for
processing streaming graphs, and provide the
full interface in Appendix~\ref{subsec:interface}. The Aspen interface is
an extension of Ligra's interface.  It includes the full Ligra
interface---\vset{}s, \emap{}, and various other functionality on a
fixed graph.  On top of Ligra, we add a set of functions for updating
the graph---in particular, for inserting or deleting sets of edges or sets of vertices.  We also add a flat-snapshot function.  Aspen currently does
not support weighted edges, but we plan to add this functionality
using a similar compression scheme for weights as used in Ligra+ in
the future. All of the functions for processing and updating the
graph work on a \emph{fixed and immutable version (snapshot)} of the graph.
The updates are functional, and therefore instead of mutating
the version, return a handle to a new graph. The implementation of
these operations follow the description given in the previous
sections.

The Aspen interface supports three functions, \acquire{}, \verset{},
and \release{}, for acquiring the current version of a graph, setting
a new version, and releasing a the version. The interface
is based on the recently defined \emph{version maintenance problem}
and implemented with the corresponding lock-free algorithm to solve
it~\cite{ben2018efficient}.  \release{} returns whether it is the last
copy on that version, and if so we garbage collect it.  The three
functions each act atomically.  The framework allows any number of
concurrent readers (i.e., transactions that \acquire{} and \release{}
but do not set) and a single writer (\acquire{}s, \verset{}s, and then
\release{}s).  Multiple concurrent readers can acquire the same
version, or different versions depending on how the writer is
interleaved with them.  The implementation of this interface is
non-trivial due to race conditions between the three operations.
Importantly, however, no reader or writer is ever blocked or delayed
by other readers or writers.  The Aspen implementation guarantees
strict serializability, which means that the state of the graph and
outputs of queries are consistent with some serial execution of the
updates and queries corresponding to real time.

Aspen is implemented in C++ and uses PAM~\cite{sun2018pam} as the
underlying purely-functional tree data structure for storing the
heads. Our \treeplus{} implementation requires about 1400 lines of
C++, most of which are for implementing \textsc{Union},
\textsc{Difference}, and \textsc{Intersect}. Our graph data structure
uses an augmented purely-functional tree from PAM to store the
vertex-tree. Each node in the vertex tree stores an
integer \treeplus{} storing the edges incident to each vertex as its
value. We note that the vertex-tree could also be compressed using a
\treeplus{}, but we did not explore this direction in the present
work. To handle memory management, our implementations use a parallel
reference counting garbage collector along with a custom pool-based
memory allocator.  The pool-allocation is critical for achieving good
performance due to the large number of small memory allocations in the
the functional setting. Although C++ might seem like an odd choice for
implementing a functional interface, it allows us to easily integrate
with PAM and Ligra. We also note that although our graph interface is
purely-functional (immutable), our global and local graph algorithms
are not. They can mutate local state within their transaction, but can
only access the shared graph through an immutable interface.

\begin{table*}
\footnotesize
\centering
\begin{tabular}[!t]{l|r|r|c}
\toprule
\textbf{Graph} & \textbf{Num. Vertices} & \textbf{Num. Edges} & \textbf{Avg. Deg.} \\
\hline
{\emph{ LiveJournal  }  }    & 4,847,571        &85,702,474      &  17.8  \\ 
{\emph{ com-Orkut    }  }    & 3,072,627        &234,370,166     &  76.2  \\ 
{\emph{ Twitter      }  }    & 41,652,231       &2,405,026,092   &  57.7  \\ 
{\emph{ ClueWeb      }  }    & 978,408,098      &74,744,358,622  &  76.4  \\ 
{\emph{ Hyperlink2014}  }    & 1,724,573,718    &124,141,874,032 &  72.0  \\ 
{\emph{ Hyperlink2012}  }    & 3,563,602,789    &225,840,663,232 &  63.3  \\ 
\end{tabular}
\captionof{table}{\small Statistics about our input graphs.
}
\label{table:sizes}
\end{table*}

\begin{table*}
\footnotesize
\centering
\begin{tabular}[!t]{l| c|c|c|c|c}
\toprule
\textbf{Graph} & \textbf{Flat Snap.} & \textbf{Aspen Uncomp.} & \textbf{Aspen (No DE)} & \textbf{Aspen (DE)} & \textbf{Savings} \\ 
\hline
{\emph{ LiveJournal  }  }    & 0.0722 & 2.77  & 0.748 & 0.582 & 4.75x \\ 
{\emph{ com-Orkut    }  }    & 0.0457 & 7.12  & 1.47  & 0.893 & 7.98x \\ 
{\emph{ Twitter      }  }    & 0.620  & 73.5  & 15.6  & 9.42  & 7.80x \\ 
{\emph{ ClueWeb      }  }    & 14.5   & 2271  & 468   & 200   & 11.3x \\ 
{\emph{ Hyperlink2014}  }    & 25.6   & 3776  & 782   & 363   & 10.4x \\ 
{\emph{ Hyperlink2012}  }    & 53.1   & 6889  & 1449  & 702   & 9.81x \\ 
\end{tabular}
\captionof{table}{\small Statistics about the
  memory usage using different formats in Aspen. \textbf{Flat Snap.}
  shows the amount of memory in GBs required to represent a flat
  snapshot of the graph. \textbf{Aspen Uncomp.}, \textbf{Aspen (No
    DE)}, and \textbf{Aspen (DE)} show the amount of memory in
  GBs required to represent the graph using uncompressed trees,
  Aspen without difference encoding of \chunks{}, and Aspen with
  difference encoding of \chunks{}, respectively.  \textbf{Savings}
  shows the factor of memory saved by using Aspen (DE) over the uncompressed representation.
}
\label{table:memory}
\end{table*}

\section{Experiments}\label{sec:exps}

\myparagraph{Algorithms} We implemented five algorithms in Aspen,
consisting of three global algorithms and two local algorithms.
Our global algorithms are breadth-first search (\defn{BFS}),
single-source betweenness centrality (\defn{BC}), and maximal
independent set (\defn{MIS}). Our BC implementation computes the
contributions to betweenness scores for shortest paths emanating from
a single vertex. The algorithms are similar to the algorithms
in~\cite{dhulipala2018theoretically} and required only minor changes
to acquire a flat snapshot and include it as an argument to \emap{}.
As argued in Section~\ref{sec:graphalgs}, the cost of creating the
snapshot does not asymptotically affect the work or depth of our
implementations. The work and depth of our
implementations of BFS, BC, and MIS are identical to the
implementations in~\cite{dhulipala2018theoretically}.
Our local algorithms are \defn{$2$-hop} and \defn{Local-Cluster}.
$2$-hop computes the set of vertices that are at most $2$ hops away
from the vertex using \emap{}. The worst-case work is $O(m + n\log n)$
and the depth is $O(\log n)$. Local-Cluster is a sequential
implementation of the Nibble-Serial graph clustering algorithm
(see~\cite{shun2016parallel,Spielman2004}), run using $\epsilon =
10^{-6}$ and $T=10$.

In our experiments, we run the global queries one at a time due to
their large memory usage and significant internal parallelism, and run
the local queries concurrently (many at the same time).

\myparagraph{Experimental Setup} Our experiments are performed on a 72-core
Dell PowerEdge R930 (with two-way hyper-threading) with $4\times 2.4\mbox{GHz}$
Intel 18-core E7-8867 v4 Xeon processors (with a 4800MHz bus and 45MB L3 cache)
and 1\mbox{TB} of main memory. Our programs use a work-stealing
scheduler that we implemented. The scheduler is implemented
similarly to Cilk for parallelism. Our programs are compiled with the
\texttt{g++} compiler (version 7.3.0) with the \texttt{-O3} flag.
All experiments involving balanced-binary trees use
weight-balanced trees as the underlying balanced tree
implementation~\cite{blelloch16justjoin, sun2018pam}.
We use Aspen to refer to the system using \treepluses{} and difference
encoding within each \chunk{} and explicitly specify other
configurations of the system if necessary.

\myparagraph{Graph Data}
Table~\ref{table:sizes} lists the graphs we use.
\defn{LiveJournal} is a directed graph of the LiveJournal social
network~\cite{boldi2004webgraph}.  \defn{com-Orkut} is an undirected
graph of the Orkut social network. \defn{Twitter} is a directed graph
of the Twitter network, where edges represent the follower
relationship~\cite{kwak2010twitter}. \defn{ClueWeb} is a Web graph
from the Lemur project at CMU~\cite{boldi2004webgraph}.
\defn{Hyperlink2012} and \defn{Hyperlink2014} are directed hyperlink
graphs obtained from the WebDataCommons dataset where nodes represent
web pages~\cite{meusel15hyperlink}. Hyperlink2012 is the \emph{largest
publicly-available graph}, and we show that \emph{Aspen is able to process it
on a single multicore machine}.  We symmetrized the graphs in our
experiments, as the running times for queries like BFS and BC are more
consistent on undirected graphs due to the majority of vertices being
in a single large component.

\myparagraph{Overview of Results}
We show the following experimental results in this section.

\begin{itemize}[topsep=0pt,itemsep=0pt,parsep=0pt,leftmargin=8pt]

\item The most memory-efficient
representation of \treepluses{} saves between 4--11x memory over using
uncompressed trees, and improves performance by 2.5--2.8x
compared to using uncompressed trees
(Section~\ref{sec:aspcompression}).

\item Algorithms implemented using Aspen are scalable, achieving between 32--78x speedup across inputs (Section~\ref{sec:scalability}).

\item Updates and queries can be run concurrently in Aspen with only a slight increase in latency (Section~\ref{sec:update-and-query}).

\item Parallel batch updates in Aspen are efficient, achieving
between 105--442M updates/sec for large batches (Section~\ref{sec:parallelupdates}).

\item Aspen outperforms \stinger{} by 1.8--10.2x while using 8.5--11.4x less memory (Section~\ref{sec:stingercomp}).

\item Aspen outperforms \llama{} by 2.8--7.8x while using 1.9--3.5x less memory (Section~\ref{sec:llamacomp}).

\item Aspen is competitive with state-of-the-art static graph processing
  systems, ranging from being 1.4x slower to 30x faster
  (Section~\ref{sec:staticexps}).

\end{itemize}

\begin{table*}
\footnotesize
\centering
\tabcolsep=0.1cm
\begin{tabular}[t]{l | c|c|c | c|c|c | c|c|c }
  \toprule
  {\bf Application} &  \multicolumn{3}{c|}{LiveJournal} & \multicolumn{3}{c|}{com-Orkut} & \multicolumn{3}{c|}{Twitter}\\
 & \textbf{(1)} & \textbf{(72h)} & \textbf{(SU)}  & \textbf{(1)} & \textbf{(72h)} & \textbf{(SU)} & \textbf{(1)} & \textbf{(72h)} & \textbf{(SU)} \\
  \midrule
  {BFS}                & 0.981 & 0.021 & 46.7       & 0.690 & 0.015 & 46.0     & 7.26  & 0.138 & 52.6       \\
  {BC}                 & 4.66 & 0.075  & 62.1       & 4.58  & 0.078  & 58.7    & 81.2  & 1.18  & 68.8       \\
  {MIS}                & 3.38 & 0.054  & 62.5       & 4.19  & 0.069  & 60.7    & 71.5  & 0.99  & 72.2       \\
  \midrule
  {$2$-hop}            & 4.36e-3 & 1.06e-4 & 41.1   & 2.95e-3 & 6.82e-5 & 43.2    & 0.036 & 8.70e-4 & 41.3  \\
  {Local-Cluster}      & 0.075   & 1.64e-3 & 45.7   & 0.122   & 2.50e-3 & 48.8    & 0.127 & 2.59e-3 & 49.0    \\
  \bottomrule
\end{tabular}
\caption{\small Running times (in seconds) of our algorithms over
  symmetric graph inputs where \textbf{(1)} is the single threaded time \textbf{(72h)} is the
  72-core time (with hyper-threading, i.e., 144 threads), and \textbf{(SU)} is the self-relative speedup.
}
\label{table:aspen-speedup}
\end{table*}

\begin{table*}
\footnotesize
\centering
\tabcolsep=0.1cm
\begin{tabular}[t]{l | c|c|c | c|c|c | c|c|c }
  \toprule
  {\bf Application} &  \multicolumn{3}{c|}{ClueWeb} & \multicolumn{3}{c|}{Hyperlink2014} & \multicolumn{3}{c|}{Hyperlink2012} \\
 & \textbf{(1)} & \textbf{(72h)} & \textbf{(SU)} & \textbf{(1)} & \textbf{(72h)} & \textbf{(SU)} & \textbf{(1)} & \textbf{(72h)} & \textbf{(SU)} \\
  \midrule
  {BFS}                & 186  & 3.69 & 50.4       & 362  & 6.19 & 58.4       & 1001 & 14.1 & 70.9   \\
  {BC}                 & 1111 & 21.8 & 50.9       & 1725 & 24.5 & 70.4       & 4581 & 58.1 & 78.8 \\
  {MIS}                & 955  & 12.1 & 78.9       & 1622 & 22.2 & 73.0       & 3923 & 50.8 & 77.2 \\
  \midrule
  {$2$-hop}            & 0.883 & 0.021   & 42.0   & 1.61   & 0.038   & 42.3    & 3.24 & 0.0755 & 42.9 \\
  {Local-Cluster}      & 0.016 & 4.45e-4 & 35.9   & 0.022  & 6.75e-4 & 32.5    & 0.028 & 6.82e-4 & 41.0 \\
  \bottomrule
\end{tabular}
\caption{\small Running times (in seconds) of our algorithms over
  symmetric graph inputs where \textbf{(1)} is the single threaded time \textbf{(72h)} is the
  72-core time (with hyper-threading, i.e., 144 threads), and \textbf{(SU)} is the self-relative speedup.
}
\label{table:aspen-speedup-2}
\end{table*}

\subsection{Chunking and Compression in Aspen}\label{sec:aspcompression}

\myparagraph{Memory Usage}
Table~\ref{table:memory} shows the amount of memory required to
represent real-world graphs in Aspen without compression, using
\treepluses{}, and finally using \treepluses{} with difference
encoding. In the uncompressed representation, the size of a vertex-tree node is 48 bytes, and the size of an edge-tree node is 32 bytes.
On the other hand, in the compressed representation, the size of a
vertex-tree node is 56 bytes (due to padding and extra pointers for
the \plusnode{}) and the size of an edge-tree node is 48 bytes. We
calculated the memory footprint of graphs that require more than 1TB
of memory in the uncompressed format by hand, using the sizes of nodes
in the uncompressed format.

We observe that by using \treepluses{} and difference encoding to
represent the edge trees, we reduce the memory footprint of the
dynamic graph representation by 4.7--11.3x compared to the
uncompressed format. Using difference encoding provides between
1.2--2.3x reduction in memory usage compared to storing the \chunks{}
in an uncompressed format.  We observe that both using \treepluses{}
and compressing within the \chunks{} is crucial for storing and
processing our largest graphs in a reasonable amount of memory.

\myparagraph{Comparison with Uncompressed Trees}
Next, we study the performance improvement gained by the improved
locality of the \treeplus{} data structure.  Due to the
memory overheads of representing large graphs using the uncompressed
format (see Table~\ref{table:memory}), we are only able to
report results for our three smallest graphs, LiveJournal, com-Orkut,
and Twitter, as we cannot store the larger graphs even with 1TB of RAM
in the uncompressed format.
We ran BFS on both the uncompressed and \treeplus{} formats (using
difference encoding) and show the results in the Appendix
(Table~\ref{table:uncomp-vs-comp}). The results show that using the
compressed representation improves the running times of these
applications from between 2.5--2.8x across these graphs.

\myparagraph{Choice of Chunk Size}
Next, we consider how Aspen performs as a function of the expected
chunk size, $b$. Table~\ref{table:chunksize-evaluation} reports the
amount of memory used, and the BFS, BC, and MIS running times as a
function of $b$. In the rest of the paper, we fixed $b=2^{8}$, which
we found gave the best tradeoff between the amount of memory consumed
(it requires 5\% more memory than the most memory-efficient
configuration) while enabling good parallelism across different
applications.

\begin{table}[!t]
\footnotesize
\centering
\tabcolsep=0.12cm
\hspace*{-0.2cm}
\begin{tabular}[t]{c | c|c|c|c }
  \toprule
  {\bf $b$ (Exp. Chunk Size)} & \textbf{Memory} & \textbf{BFS (72h)} & \textbf{BC (72h)} & \textbf{MIS (72h)}\\
  \midrule
  {$2^{1}$}           & 68.83 & 0.309    & 2.72  &  2.17 \\ 
  {$2^{2}$}           & 41.72 & 0.245    & 2.09  &  1.71 \\ 
  {$2^{3}$}           & 26.0 & 0.217     & 1.68  &  1.41 \\ 
  {$2^{4}$}           & 17.7 & 0.172     & 1.45  &  1.24 \\ 
  {$2^{5}$}           & 13.3 & 0.162     & 1.32  &  1.14 \\ 
  {$2^{6}$}           & 11.1 & 0.152     & 1.25  &  1.07 \\ 
  {$2^{7}$}           & 9.97 & 0.142     & 1.22  &  1.01 \\ 
{$\mathbf{2^{8}}$} &  9.42 & {\bf 0.138} & {\bf 1.18}  &  0.99 \\ 
  {$2^{9}$}           & 9.17 & 0.141     & 1.20  &  0.99 \\ 
  {$2^{10}$}          & 9.03 & 0.152     & 1.19  &  0.98 \\
  {$2^{11}$}          & 8.96 & 0.163     & 1.20  &  0.98 \\
  {$2^{12}$}          & {\bf 8.89} & 0.170     & 1.21  &  {\bf 0.98} \\
  \bottomrule
\end{tabular}
\caption{Memory usage (gigabytes) and performance (seconds) for the Twitter
  graph as a function of the (expected) chunk size. All times are measured on 72
  cores using hyper-threading.   Bold-text marks the best value in each  
  column.   We use $2^{8}$ in the other experiments.
}
\label{table:chunksize-evaluation}. 
\end{table}

\subsection{Parallel Scalability of Aspen}\label{sec:scalability}

\myparagraph{Algorithm Performance}
Tables~\ref{table:aspen-speedup} and \ref{table:aspen-speedup-2} report experimental results including
the single-threaded time and 72-core time (with hyper-threading) for
Aspen using compressed \treepluses{}.
For BFS, we achieve
between 46--70x speedup across all inputs. For BC, our implementations
achieve between 50--78x speedup across all inputs. Finally, for MIS, our implementations achieve between
60x--78x speedup across all inputs.
We observe that the experiments
in~\cite{dhulipala2018theoretically} report similar speedups for the
same graphs.
For local algorithms, we report
the average running time for performing 2048 queries sequentially and
in parallel. We achieve between 41--43x speedup for $2$-hop, and between
35--49x speedup for Local-Cluster.

\begin{table}[!t]
\footnotesize
\centering
\tabcolsep=0.12cm
\hspace*{-0.2cm}
\begin{tabular}[t]{l | c|c|c | c}
  \toprule
  {\bf Graph} & \textbf{Without FS} & \textbf{With FS} &
  \textbf{Speedup} & \textbf{FS Time} \\
  \midrule
  {LiveJournal}        & 0.028 & 0.021  & 1.33 & 3.8e-3  \\ 
  {com-Orkut}          & 0.018 & 0.015  & 1.12 & 2.3e-3  \\
  {Twitter}            & 0.184 & 0.138  & 1.33 & 0.034   \\
  {ClueWeb}            & 4.98  & 3.69   & 1.34 & 0.779   \\
  {Hyperlink2014}      & 7.51  & 6.19   & 1.21 & 1.45   \\
  {Hyperlink2012}      & 18.3  & 14.1   & 1.29 & 3.03   \\
  \bottomrule
\end{tabular}
\caption{72-core with hyper-threading running times (in seconds) comparing the performance
  of BFS without flat snapshots (\textbf{Without FS}) and with flat snapshots
  (\textbf{With FS}), as well as the running time for computing the
  flat snapshot (\textbf{FS Time}).}
\label{table:flat-snapshot-comparison}
\end{table}

\myparagraph{Flat Snapshots}
Table~\ref{table:flat-snapshot-comparison} shows the running times of
BFS with and without the use of a flat snapshot. Our BFS implementation is
between 1.12--1.34x faster using a flat snapshot, including the time to compute
a flat snapshot. The table also
reports the time to acquire a flat snapshot, which is between 15--24\% of the
overall BFS time across all graphs.
We observe that acquiring a flat snapshot is already an improvement for a single
run of an algorithm, and quickly becomes more profitable as multiple algorithms
are run over a single snapshot of the graph (e.g., multiple
BFS's or betweenness centrality computations).

\subsection{Simultaneous Updates and Queries}\label{sec:update-and-query}
In this sub-section, we experimentally verify that Aspen can support
low-latency queries and updates running concurrently. In these experiments, we
generate an update stream by randomly sampling 2 million edges from
the input graph to use as updates. We sub-sample 90\% of the sample to
use as edge insertions, and immediately delete them from the input
graph. The remaining 10\% are kept in the graph, as we will delete them
over the course of the update stream. The update stream is a random
permutation of these insertions and deletions. We believe that
sampling edges from the input graph better preserves the properties of
the graph and ensures that edge deletions perform non-trivial work,
compared to using random edge updates.

\begin{table}[!t]
\footnotesize
\centering
\tabcolsep=0.12cm
\hspace*{-0.2cm}
\begin{tabular}[t]{l | c|c | c|c}
  \toprule
  {\bf Graph} &  \multicolumn{2}{c|}{\textbf{Update}} & \multicolumn{2}{c}{\textbf{Query (BFS)}}\\
  & \textbf{Edges/sec} & \textbf{Latency} & \textbf{Latency (C)} & \textbf{Latency (I)} \\
  \midrule
  {LiveJournal}   & 7.86e4  & 1.27e-5 & 0.0190 & 0.0185 \\ 
  {com-Orkut}     & 6.02e4  & 1.66e-5 & 0.0179 & 0.0176 \\
  {Twitter}       & 4.44e4  & 1.73e-5 & 0.155  & 0.155 \\
  {ClueWeb}       & 2.06e4  & 4.83e-5 & 4.83   & 4.82 \\
  {Hyperlink2014} & 1.42e4  & 7.04e-5 & 6.17   & 6.15 \\
  {Hyperlink2012} & 1.16e4  & 8.57e-5 & 15.8   & 15.5 \\
  \bottomrule
\end{tabular}
\caption{\small Throughput and average latency achieved by Aspen when
  concurrently processing a sequential stream of edge updates along with a
  sequential stream of breadth-first search queries (each BFS is internally
  parallel). \textbf{Latency (C)} reports the average latency of the query when
  running the updates and queries concurrently, while \textbf{Latency (I)}
  reports the average latency when running queries in isolation on the modified
  graph.}
\label{table:sequential-throughput}
\end{table}

After constructing the update stream, we spawn two parallel jobs, one
which performs the updates sequentially and one which performs global
queries. We maintain the undirectedness of the graph by inserting each
edge as two directed edge updates, within a single batch. For global queries, we
run a stream of BFS's from random sources one after the other and measure the
average latency. We note that for the BFS queries, as our inputs are
symmetrized, a random vertex is likely to fall in the giant connected component
which exists in all of our input graphs. The global queries therefore process
nearly all of the vertices and edges.

Table~\ref{table:sequential-throughput} shows the throughput in terms
of directed edge updates per second, the average latency to make an undirected
edge visible, and the latency of global queries both when running concurrently
with updates and when running in isolation. We note that when running global
queries in isolation, we use all of the threads in the system (72-cores with
hyper-threading). We observe that our data structure achieves between 22--157
thousand directed edge updates per second, which is achieved while concurrently
running a parallel query on all remaining threads.  We obtain higher update
rates on smaller graphs, where the small size of the graph enables it to utilize
the caches better. In all cases, the average latency for making an edge visible
is at most 86 microseconds, and is as low as 12.7 microseconds on the smallest
graph.

The last two columns in Table~\ref{table:sequential-throughput} show
the average latency of BFS queries from random sources when running
queries concurrently with updates, and when running queries in
isolation. We see that the performance impact of running updates
concurrently with queries is less than 3\%, which could be due to having one
fewer thread.  We ran a similar experiment, where we ran updates on 1 core and
ran multiple concurrent local queries (Local-Cluster) on the remaining cores,
and found that the difference in average query times is even lower than for BFS.




\subsection{Performance of Batch Updates}\label{sec:parallelupdates}
In this sub-section, we show that the batch versions of our primitives
achieve high throughput when updating the graph, even on very large
graphs and for very large batches. As there are insufficient edges on
our smaller graphs for applying the methodology from
Section~\ref{sec:update-and-query}, we sample directed edges from an rMAT
generator~\cite{chakrabarti2004r} with $a=0.5, b=c=0.1, d=0.3$ to
perform the updates. To evaluate our performance on a batch of size
$B$, we generate $B$ directed edge updates from the stream (note that there can
be duplicates), repeatedly call \insertedges{} and \deleteedges{} on the batch,
and report the median of three such trials. The costs that we report
\emph{include} the time to sort the batch and combine duplicates.


\begin{table}[!t]
\footnotesize
\centering
\tabcolsep=0.12cm
\hspace*{-0.2cm}
\begin{tabular}[t]{l | c|c|c|c|c|c}
  \toprule
  {\bf Graph} &  \multicolumn{6}{c}{\bf Batch Size}\\
  & $10$ & $10^{3}$ & $10^{5}$ & $10^{7}$ & $10^{9}$ & $2\cdot 10^{9}$\\
  \midrule
  {LiveJournal}   &8.26e4  & 2.88e6 & 2.29e7 & 1.56e8 & 4.13e8 & 4.31e8 \\ 
  {com-Orkut}     &7.14e4  & 2.79e6 & 2.22e7 & 1.51e8 & 4.21e8 & 4.42e8 \\
  {Twitter}       &6.32e4  & 2.63e6 & 1.23e7 & 5.68e7 & 3.04e8 & 3.15e8 \\
  {ClueWeb}       &6.57e4  & 2.38e6 & 7.19e6 & 2.64e7 & 1.33e8 & 1.69e8 \\
  {Hyperlink2014} &6.17e4  & 2.12e6 & 6.66e6 & 2.28e7 & 9.90e7 & 1.39e8 \\
  {Hyperlink2012} &6.45e4  & 2.04e6 & 4.97e6 & 1.84e7 & 8.26e7 & 1.05e8 \\
  \bottomrule
\end{tabular}
\caption{Throughput (directed edges/second) obtained when performing parallel
  batch edge insertions on different graphs with varying batch sizes, where
  inserted edges are sampled from an rMAT graph generator. We note that the
  times for batch deletions are similar to the time for insertions. All times
  are on 72 cores with hyper-threading.
}
\label{table:parallel-throughput}
\end{table}

\begin{figure}[!t]
  \centering
  \includegraphics[width=0.75\textwidth]{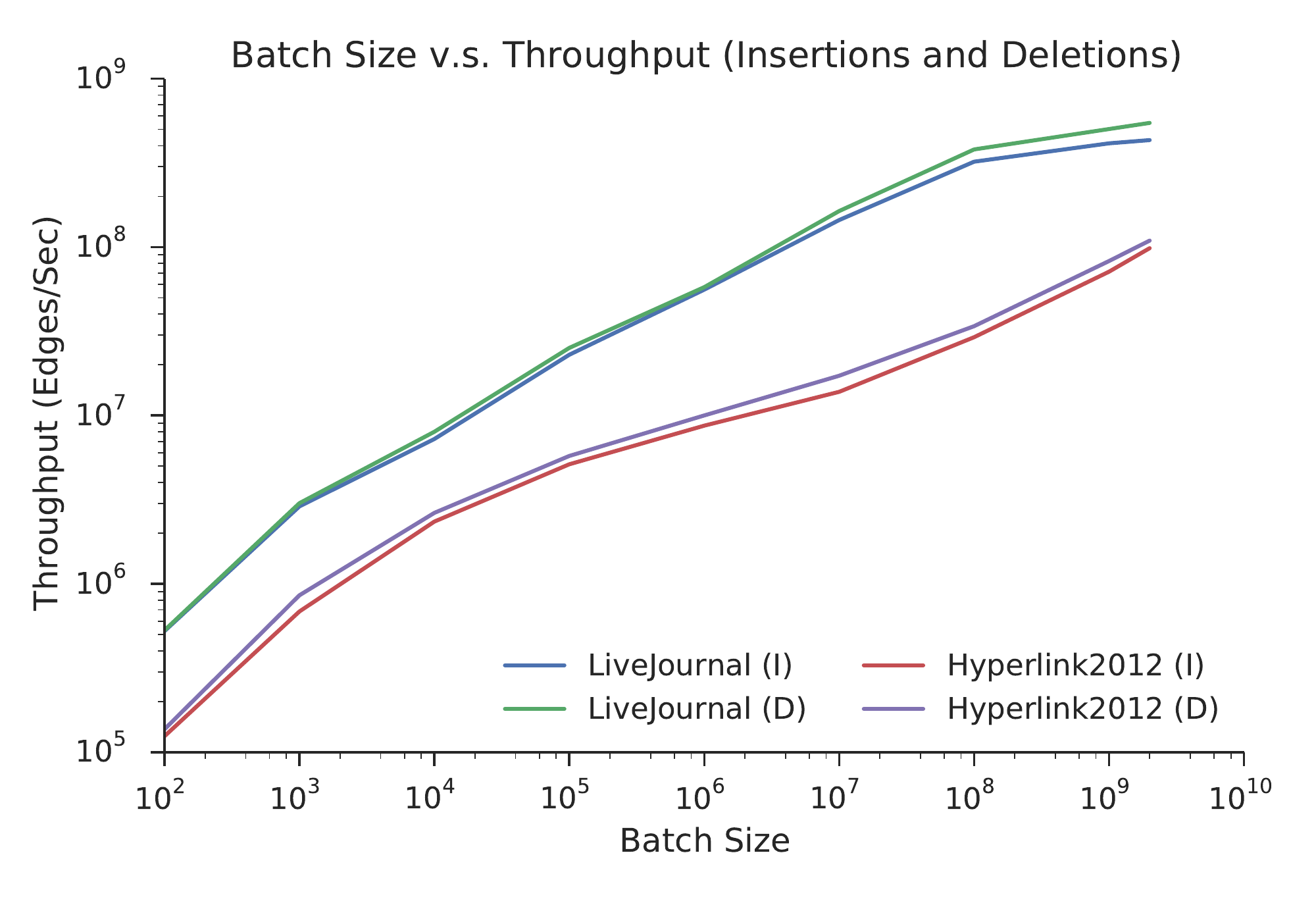}
    \caption{Throughput (edges/sec) when performing batches of insertions (I)
      and deletions (D) with varying batch sizes on Hyperlink2012 and
      LiveJournal in a log-log scale.  All times are on 72 cores with
      hyper-threading.}\label{fig:par-throughput}
\end{figure}

Table~\ref{table:parallel-throughput} shows the throughput (the number of edges
processed per second) of performing batch edge insertions in parallel on varying
batch sizes.  The throughput for edge deletions are within 10\% of the edge
insertion times, and are usually faster (see Figure~\ref{fig:par-throughput}).
The running time can be calculated by dividing the batch size by the throughput.
We illustrate the throughput obtained for both insertions and deletions in
Figure~\ref{fig:par-throughput} for the largest and smallest graph, and note
that the lines for other graphs are sandwiched between these two lines. The
only exception of com-Orkut, where batch insertions achieve about 2\% higher
throughput than soc-LiveJournal at the two largest batch sizes.

We observe that Aspen's throughput seems to vary depending on the
graph size. We achieve a maximum throughput of 442M updates per second
on com-Orkut when processing batches of 2B updates. On the other hand,
on the Hyperlink2012 graph, the largest graph that we tested on, we
achieve 105M updates per second for this batch size. We believe that
the primary reason that small graphs achieve much better throughput at
the largest batch size is that nearly all of the vertices in the tree
are updated for the small graphs. In this case, due to the asymptotic
work bound for the update algorithm, the work for our updates become
essentially linear in the tree size.

\subsection{Comparison with \stinger{}}\label{sec:stingercomp}

In this sub-section, we compare Aspen to
Stinger~\cite{ediger2012stinger}, a state-of-the-art graph-streaming
system.

\myparagraph{\stinger{} Design}
\stinger{}'s data structure for processing streaming graphs is based on
adapting the CSR format to support dynamic updates. Instead of storing
all edges of a vertex contiguously, it chunks the edges into a number
of blocks, which are chained together as a linked list. Updates
traverse the list to find an empty slot for a new edge, or to
determine whether an edge exists. Therefore, updates take $O(deg(v))$ work
and depth for a vertex $v$ that is updated. Furthermore, updates use
fine-grained locking to perform edge insertions, which may
result in contention when updating very high degree vertices.  As
\stinger{} does not support compressed graph inputs, we were unable to
run the system on our input graphs that are larger than Twitter.

\myparagraph{Memory Usage}
We list the sizes of the three graphs that \stinger{} was able to process in
Table~\ref{table:stinger-llama-size}. The \stinger{} interface supports a
function which returns the size of its in-memory representation in
bytes, which is what we use to report the numbers in this paper.

We found that \stinger{} has a high memory usage, even in the
memory-efficient settings used in our experiments. The memory usage we
observed appears to be consistent with~\cite{ediger2012stinger}, which
reports that the system requires 313GB of memory to store a scale-free
(RMAT) graph with 268 million vertices and 2.15 billion edges, making
the cost 145 bytes per edge. This number is on the same order of
magnitude as the numbers we report in
Table~\ref{table:stinger-llama-size}. We found that Aspen is between
8.5--11.4x more memory efficient than \stinger{}.


{
\setlength{\tabcolsep}{2pt}
\begin{table}[!t]
\footnotesize
\centering
\begin{tabular}[!t]{l|c|c|c|c|c|c|c}
\toprule
\textbf{Graph} & \textbf{ST} & \textbf{LL} &\textbf{Ligra+} &
\textbf{Aspen} & \textbf{ST/Asp.} & \textbf{LL/Asp.} & \textbf{L+/Asp.} \\
\hline
{\emph{ LiveJournal   }  }  & 4.98  & 1.12  & 0.246  & 0.582 & 8.55x & 1.92x & 0.422x \\
{\emph{ com-Orkut     }  }  & 10.2  & 3.13  & 0.497  & 0.893 & 11.4x & 3.5x & 0.55x \\
{\emph{ Twitter       }  }  & 81.8  & 31.4  & 5.1    & 9.42  & 8.6x  & 3.3x & 0.54x \\
{\emph{ ClueWeb       }  }  & --    & --    & 100    & 200   & --    & --   & 0.50x \\
{\emph{ Hyperlink2014 }  }  & --    & --    & 184    & 363   & --    & --   & 0.50x \\
{\emph{ Hyperlink2012 }  }  & --    & --    & 351    & 702   & --    & --   & 0.50x \\

\end{tabular}
\captionof{table}{\small
The first four columns show the memory in gigabytes required to
represent the graph using \stinger{} (\textbf{ST}), \llama{} (\textbf{LL}),
Ligra+, and Aspen respectively.  \textbf{ST/A}, \textbf{LL/A}, and
\textbf{L+/A} is the amount of memory used by \stinger{}, \llama{}, and
Ligra+ divided by the memory used by Aspen respectively. \stinger{} and
\llama{} do not support compression and were not able to store the
largest graphs used in our experiments. }
\label{table:stinger-llama-size}
\end{table}
}

\myparagraph{Batch Update Performance}
We measure the batch update performance of \stinger{} by using an rMAT
generator provided in \stinger{} to generate the directed updates. We set $n =
2^{30}$ for updates in the stream.  The largest batch size supported by
\stinger{} is 2M directed updates.
  The update times for \stinger{} were
fastest when inserting into nearly-empty graphs. For each batch size, we insert
10 batches of edges of that size into the graph, and report the median time.

The results in Table~\ref{table:stinger-throughput} show the update
rates for inserting directed edge updates in \stinger{} and Aspen. We observe
that the running time for \stinger{} is reasonably high, even on very small
batches, and grows linearly with the size of the batch.  The Aspen update times
also grow linearly, but are very fast for small batches.  Perhaps surprisingly,
our update time on a batch of 1M updates is faster than the update time of
\stinger{} on a batch of 10 edges.

\begin{table}[!t]
\footnotesize
\centering
\tabcolsep=0.12cm
\hspace*{-0.2cm}
\begin{tabular}[t]{c | c|c|c|c}
  \toprule
  {\bf Batch Size} & \textbf{\stinger{}} & \textbf{Updates/sec} & \textbf{Aspen} & \textbf{Updates/sec} \\
  \midrule
  10             & 0.0232 & 431     & 9.74e-5 & 102,669  \\ 
  $10^{2}$       & 0.0262 & 3,816   & 2.49e-4 & 401,606 \\
  $10^{3}$       & 0.0363 & 27,548  & 6.98e-4 & 1.43M \\
  $10^{4}$       & 0.171  & 58,479  & 2.01e-3 & 4.97M \\
  $10^{5}$       & 0.497  & 201,207 & 9.53e-3 & 10.4M \\
  $10^{6}$       & 3.31   & 302,114 & 0.0226  & 44.2M \\
  $2\cdot10^{6}$ & 6.27   & 318,979 & 0.0279  & 71.6M \\
  \bottomrule
\end{tabular}
\caption{Running times and update rates (directed edges/second) for \stinger{}
  and Aspen when performing batch edge updates on an empty graph with varying
  batch sizes. Inserted edges are sampled from the RMAT graph generator. All
  times are on 72 cores with hyper-threading.
}
\label{table:stinger-throughput}

\end{table}

\myparagraph{Algorithm Performance} Lastly, we show the performance of
graph algorithms implemented using the \stinger{} data structures. We
use the BFS implementation for \stinger{} developed in McColl et
al.~\cite{mccoll2014performance}.
We used a BC implementation that is available in the \stinger{} code base.
Unfortunately, this implementation is entirely sequential, and so we
compare \stinger{}'s BC time to our single-threaded time.
Neither of the \stinger{} implementations perform
direction-optimization, so to perform a fair comparison, we used an
implementation of BFS and BC in Aspen that disables direction-optimization.
Table~\ref{table:stinger-llama-comparison} shows the parallel running
times of of \stinger{} and Aspen for these problems. For BFS, which is
run in parallel, we achieve between 6.7--10.2x speedup over \stinger{}.
For BC, which is run sequentially, we achieve between 1.8--4.2x speedup
over \stinger{}. A likely reason that Aspen's BFS is significantly faster
than \stinger{}'s is that it can process edges incident to high-degree
vertices in parallel, whereas traversing a vertex's neighbors in
\stinger{} requires sequentially traversing a linked list of blocks.


\begin{table}[!t]
\footnotesize
\centering
\tabcolsep=0.12cm
\hspace*{-0.2cm}
\begin{tabular}[t]{l| l | c|c|c|c|c | c|c }
  \toprule
  {\bf App.} & {\bf Graph}
	& \textbf{ST} & \textbf{LL} & \textbf{A} & \textbf{A(1)} & \textbf{A}$^{\dagger}$ & \textbf{ST/A} & \textbf{LL/A} \\
  \midrule
 \parbox[t]{2mm}{\multirow{3}{*}{BFS }}
 & {LiveJournal}  & 0.478 & 0.161 & 0.047 & -- & 0.021 & 10.2 & 3.42 \\ 
 & {com-Orkut}    & 0.548 & 0.192 & 0.067 & -- & 0.015 & 8.18 & 2.86 \\
 & {Twitter}      & 6.99  & 8.09  & 1.03  & -- & 0.138 & 6.79 & 7.85 \\

  \midrule
 \parbox[t]{2mm}{\multirow{3}{*}{BC }}

 & {LiveJournal}  & 18.7 & 0.408 & 0.105 & 5.45 & 0.075 & 3.43 & 3.88 \\ 
 & {com-Orkut}    & 32.8 & 1.32  & 0.160 & 7.74 & 0.078 & 4.23 & 8.25 \\
 & {Twitter}      & 223  & 53.1  & 3.52  & 122  & 1.18  & 1.82 & 15.1 \\
\bottomrule
\end{tabular}
\caption{Running times (in seconds) comparing the performance of algorithms
  implemented in \stinger{} (\textbf{ST}), \llama{} (\textbf{LL}), and Aspen.
  \textbf{A} is the parallel time using Aspen \emph{without direction-optimization}.
  \textbf{(A(1))} is the one-thread time of Aspen, which is only relevant for comparing with \stinger{}'s BC implementation.
  \textbf{A}$^{\mathbf{\dagger}}$ is the parallel time using Aspen \emph{with direction-optimization}.
  \textbf{(ST/A)} is Aspen's speedup over \stinger{} and \textbf{(LL/A)} is Aspen's speedup over \llama{}.
}
\label{table:stinger-llama-comparison}
\end{table}

\subsection{Comparison with \llama{}}\label{sec:llamacomp}

In this sub-section, we compare Aspen to \llama{}~\cite{macko2015llama},
another state-of-the-art graph-streaming system.

\myparagraph{\llama{} Design}
Like \stinger{}, \llama{}'s streaming graph data structure is motivated
by the CSR format. However, like Aspen, \llama{} is designed for
batch-processing in the single-writer multi-reader setting and can
provide serializable snapshots. In \llama{}, a batch of size $k$ generates
a new snapshot which uses $O(n)$ space to store a vertex array, and
$O(k)$ space to store edge updates in a dynamic CSR structure. The
structure creates a linked list over the edges incident to a vertex
that is linked over multiple snapshots. This design can cause the
depth of iterating over the neighbors of a vertex to be large if the
edges are spread over multiple snapshots.

Unfortunately, the publicly-available code for \llama{} does not provide
support for evaluating streaming graph algorithms or batch updates.
However, we we were able to load static graphs and run several
implementations of algorithms in \llama{} for which we report times in
this section. As \llama{} does not support compressed graph inputs, we
were unable to run the system on our input graphs that are larger than
Twitter.

\myparagraph{Memory Usage}
Unfortunately, we were not able to get \llama{}'s internal allocator to
report correct memory usage statistics for its internal allocations.
Instead, we measured the lifetime memory usage of the process and use
this as an estimate for the size of the in-memory data structure built
by \llama{}.  The memory usage in bytes for the three graphs that \llama{}
was able to process is shown in Table~\ref{table:stinger-llama-size}.
The cost in terms of bytes/edge for \llama{} appears to be consistent,
which matches the fact that the internal representation is a flat CSR,
since there is a single snapshot. Overall, Aspen is between 1.9--3.5x
more memory efficient than \llama{}.

\myparagraph{Algorithm Performance}
We measured the performance of a parallel breadth-first search (BFS)
and single-source betweenness centrality (BC) algorithms in \llama{}. The
same source is used for both \llama{} and Aspen for both BFS and BC.
BFS and BC in \llama{} do not use direction-optimization, and so we
report our times for these algorithms without using
direction-optimization to ensure a fair comparison.

Table~\ref{table:stinger-llama-comparison} shows the running times for
BFS and BC. We achieve between 2.8--7.8x speedup over \llama{} for BFS and
between 3.8--15.1x
speedup over \llama{} for BC. \llama{}'s poor performance on
these graphs, especially Twitter, is likely due to sequentially
exploring the out-edges of a vertex in the search, which is slow on
graphs with high degrees.


\subsection{Static Graph Processing Systems}\label{sec:staticexps}

We compared Aspen to Ligra+, a state-of-the-art shared-memory graph
processing system, GAP, a state-of-the-art graph processing
benchmark~\cite{BeamerAP15}, and Galois, a shared-memory parallel programming
library for C++~\cite{Nguyen2013}.


\myparagraph{Ligra+}
Table~\ref{table:gap-galois-comparison}
the parallel running times of our three global algorithms expressed using Aspen and
Ligra+. The results show that Ligra is 1.43x faster than Aspen for
global algorithms on our small inputs.
We also performed a more extensive experimental comparison between
Aspen and Ligra+, comparing the parallel running times of all of our
algorithms on all of our inputs (Tables~\ref{table:aspen-vs-ligra-small} and
\ref{table:aspen-vs-ligra-large}).
Compared to Ligra+, across all inputs, algorithms in Aspen are 1.51x
slower on average (between 1.2x--1.7x)
for the global algorithms, and
1.45x slower on average (between 1.0--2.1x) for the local algorithms. We report
the local times in Tables~\ref{table:aspen-vs-ligra-small} and
\ref{table:aspen-vs-ligra-large}.
The local algorithms have a modest slowdown compared to their Ligra+
counterparts, due to logarithmic work vertex accesses being amortized
against the relative high average degrees (see
Table~\ref{table:sizes}).


\begin{table}[!t]
\footnotesize
\centering
\tabcolsep=0.12cm
\hspace*{-0.2cm}
\begin{tabular}[t]{l| l | c|c|c|c | c|c|c }
  \toprule
  {\bf App.} & {\bf Graph}
  & \textbf{GAP} & \textbf{Galois} & \textbf{Ligra+} & \textbf{Aspen} & {$\frac{\mathbf{GAP}}{\mathbf{A}}$} &  {$\frac{\mathbf{GAL}}{\mathbf{A}}$} & {$\frac{\mathbf{L+}}{\mathbf{A}}$} \\
  \midrule
 \parbox[t]{2mm}{\multirow{3}{*}{BFS }}
  & {LiveJ}     & 0.0238  & 0.0761 & 0.015 & 0.021  & 1.1x & 3.6x & 0.71x \\ 
  & {Orkut}     & 0.0180  & 0.0661 & 0.012 & 0.015  & 1.2x & 4.4x & 0.80x \\
  & {Twitter}   & 0.139   & 0.461  & 0.081 & 0.138  & 1.0x & 3.3x & 0.58x \\

  \midrule
 \parbox[t]{2mm}{\multirow{3}{*}{BC }}
  & {LiveJ}     & 0.0930 &  --  & 0.052 & 0.075  & 1.24x  & -- & 0.69x \\ 
  & {Orkut}     & 0.107  &  --  & 0.062 & 0.078  & 1.72x  & -- & 0.79x \\
  & {Twitter}   & 2.62   &  --  & 0.937 & 1.18   & 2.22x  & -- & 0.79x \\

  \midrule
 \parbox[t]{2mm}{\multirow{3}{*}{MIS}}
  & {LiveJ}     & -- & 1.65 & 0.032 & 0.054 & --  & 30x   & 0.59x \\ 
  & {Orkut}     & -- & 1.52 & 0.044 & 0.069 & --  & 22x   & 0.63x \\
  & {Twitter}   & -- & 8.92 & 0.704 & 0.99  & --  & 9.0x  & 0.71x \\

  \bottomrule
\end{tabular}
\caption{Running times (in seconds) comparing the performance of algorithms
  implemented in GAP, Galois, Ligra+, and Aspen.
  {$\frac{\mathbf{GAP}}{\mathbf{A}}$},
  {$\frac{\mathbf{GAL}}{\mathbf{A}}$}, and
  {$\frac{\mathbf{L+}}{\mathbf{A}}$}
  are Aspen's speedups over GAP, Galois, and Ligra+ respectively.
}
\label{table:gap-galois-comparison}
\end{table}

\myparagraph{GAP}
Table~\ref{table:gap-galois-comparison} shows the parallel running
times of the BFS and BC implementations from GAP. On average, our
implementations in Aspen are 1.4x faster than the implementations from
GAP over all problems and graphs.  We note that the code in GAP has been
hand-optimized using OpenMP scheduling primitives.  As a result, the GAP code is
significantly more complex than our code, which only uses the high-level
primitives defined by Ligra+.

\myparagraph{Galois}
Table~\ref{table:gap-galois-comparison} shows the running times of
using Galois, a shared-memory parallel programming library that
provides support for graph processing~\cite{Nguyen2013}.
Galois' algorithms (e.g., for BFS and MIS) come with several versions.
In our experiments, we tried all versions of their algorithms, and
report times for the fastest one. On average, our implementations in
Aspen are 12x faster than Galois. For BFS, Aspen is between 3.3--4.4x
faster than Galois. We note that the Galois BFS implementation is
synchronous, and does not appear to use Beamer's
direction-optimization.
We omit BC as we were not able to obtain reasonable numbers on our inputs using
their publicly-available code (the numbers we obtained were much worse than the
ones reported in~\cite{Nguyen2013}).
For MIS, our
implementations are between 9--30x faster than Galois.



\section{Related Work}\label{sec:relwork}

We have mentioned some other schemes for chunking in
Section~\ref{subsec:otherapproach}.  Although we use functional trees
to support snapshots, many other systems for supporting persistence and
snapshots use version
lists~\cite{becker1996asymptotically,reed78,persistent}. The idea is
for each mutable value or pointer to keep a timestamped list of
versions, and reading a structure to go through the list to find the
right one (typically the most current is kept first).
LLAMA~\cite{macko2015llama} uses a variation of this idea.  However,
it seems challenging to achieve the low space that we achieve using
such systems since the space for such a list is large.

\subsection{Graph Processing Frameworks}

Many
processing frameworks have been designed to process static graphs (e.g.
\cite{Dathathri2018,PrountzosMP15,Pai2016,WangXSL15,malewicz10pregel,gonzalez2012powergraph,low2010graphlab,Nguyen2013,shun2012ligra},
among many others).
We refer the reader
to~\cite{McCune2015,Yan2017} for surveys of existing frameworks.
Similar to Ligra+~\cite{shun2015ligraplus}, Log(Graph)~\cite{Besta2018} supports running parallel algorithms on compressed graphs.
Their experiments show that they have a moderate performance slowdown on real-world graphs, but sometimes get improved performance on synthetic graphs~\cite{Besta2018}. 



Existing dynamic graph streaming frameworks can be divided into two
categories based on their approach to ingesting updates. The first
category processes updates and queries in phases, i.e., updates wait
for queries to finish before updating the graph, and queries wait for
updates to finish before viewing the graph. Most existing systems take
this approach, as it allows updates to mutate the underlying graph
without worrying about the consistency of
queries~\cite{ediger2012stinger, feng2015distinger,
  green2016custinger,winter2017autonomous,Ammar2018,Sha2017,Sengupta2016,Sengupta17,Murray2016,Cai2012,Vora2017,Suzumura2014,Shi2016,Hornet}. Hornet~\cite{Hornet},
one of the most recent systems in this category, reports a throughput
of up to 800 million edges per second on a GPU with 3,840 cores (about
twice our throughput using 72 CPU cores for similarly-sized graphs);
however the graphs used in Hornet are much smaller that what Aspen can
handle due to memory limitations of GPUs.  The second category enables
queries and updates to run concurrently by isolating queries to run on
snapshots and periodically have updates generate new
snapshots~\cite{cheng2012kineograph,macko2015llama,Iyer2016,iyer2015cellq}.

GraphOne~\cite{Kumar2019} is a system developed concurrently with our
work that can handle queries running on the most recent version of the
graph while updates are running concurrently by using a combination of
an adjacency list and an edge list. They report an update rate of
about 66.4 million edges per second on a Twitter graph with 2B edges
using 28 cores; Aspen is able to ingest 94.5 million edges per second
on a larger Twitter graph using 28 cores. However, GraphOne also backs
up the update data to disk for durability.


There are also many systems that have been built for analyzing graphs
over time~\cite{khurana2013efficient, han2014chronos, KhuranaD16,
  miao2015immortalgraph, michail2016introduction,
  hartmann2017analyzing,
  fouquet2018enabling,RenLKZC11,Then2017,Vora2016}. These systems are
similar to processing dynamic graph streams in that updates to the
graph must become visible to new queries, but are different in that
queries can performed on the graph as it appeared at any point in
time. Although we do not explore historical queries in this paper,
functional data structures are particularly well-suited for this
scenario since it is easy to keep any number of persistent versions
simply by keeping their roots.

\subsection{Graph Databases}
There has been significant research on graph databases
(e.g.,~\cite{Bronson2013,Khandelwal2017,Shao2013,Prabhakaran2012,Kumar2016,dubey2016weaver,Neo4j}).
The main difference between processing dynamic graph-streams and graph databases
is that graph databases support transactions, i.e., multi-writer
concurrency. A graph database running with snapshot isolation
could be used to solve the same problem we solve.  However, due to
their need to support transactions, graph databases have significant
overhead even for graph analytic queries such as PageRank and shortest
paths.  McColl et al.~\cite{mccoll2014performance} show that Stinger
is orders of magnitude faster than 
state-of-the-art graph databases.

\section{Conclusion}
We have presented a compressed fully-functional tree data structured
called the \treeplus{} that has theoretically-efficient operations,
low space usage, and good cache locality.  We use \treeplus{}s to
represent graphs, and design a graph-streaming framework called Aspen
that is able to support concurrent queries and updates to the graph
with low latency.  Experiments show that Aspen outperforms
state-of-the-art graph-streaming frameworks, Stinger and LLAMA, and
only incurs a modest overhead over state-of-the-art static graph processing
frameworks.  Future work includes designing incremental graph algorithms and
historical queries using Aspen, and using \treeplus{}s in other applications.
Although our original motivation for
designing \treepluses{} was for representing compressed graphs, we
believe that they are of independent interest and can be used in
applications where ordered sets of integers are dynamically
maintained, such as compressed inverted indices in search engines.

\section*{Acknowledgements}
This research was supported in part by
NSF grants \#CCF-1408940, \#CF-1533858, \#CCF-1629444,  and \#CCF-1845763, and DOE grant
\#DE-SC0018947.
We thank the reviewers for their helpful comments.

\bibliographystyle{abbrv}
\bibliography{ref}

\section{Appendix}

\subsection{Parallel Primitives}
The following parallel procedures are used to describe our algorithms.
\defn{Scan} takes as input an array $A$ of length $n$, an associative
binary operator $\oplus$, and an identity element $\bot$ such that
$\bot \oplus x = x$ for any $x$, and returns the array
$(\bot, \bot \oplus A[0], \bot \oplus A[0] \oplus A[1], \ldots, \bot \oplus_{i=0}^{n-2} A[i])$
as well as the overall sum, $\bot \oplus_{i=0}^{n-1} A[i]$.
Scan can be done in $O(n)$ work and $O(\log n)$ depth (assuming $\oplus$
takes $O(1)$ work)~\cite{JaJa92}.
\defn{Filter} takes an array $A$ and a predicate $f$ and returns a new array
containing $a \in A$ for which $f(a)$ is $\mathsf{true}$, in the same order as in $A$.
Filter can be done in $O(n)$ work and $O(\log n)$ depth (assuming $f$ takes $O(1)$ work).

\subsection{Details on \treeplus{} Properties}\label{subsec:properties}

We now provide the proof for Lemma~\ref{lem:structure}.

\begin{proof}
Each \element{} is selected as a head with probability $1/b$, and so by
linearity of expectations, the expected number of heads is $n/b$.
Define $X_i$ to
be the independent random variable that is $1$ if $E_i$ is a head and
$0$ otherwise. Let $X$ be their sum, and  $E[X] = n/b$.
Applying a Chernoff bound
proves that the number of heads is $O(n/b)$ w.h.p.

We now show that each tail is not too large w.h.p. Consider a
subsequence of length $t = b \cdot (c \ln n)$ for a constant $c>1$. The probability that
none of the $t$ \elements{} in the subsequence are selected as a head is $(1 - 1/b)^t \le (1/e)^{c\ln n} = 1/n^c$.
Therefore, a subsequence of $E$ of length $t$ has a head w.h.p. We
complete the proof by applying a union bound over all length $t$
subsequences of $E$.
\end{proof}

\subsection{Details on \treeplus{} Primitives}\label{subsec:primdetails}

This section provides details missing from the main body of the paper
on how to build, search, and map over the \treeplus{} data structure.

\myparagraph{Building}
Building (Build$(S, f_{V})$) the data structure can be done in
$O(n\log n)$ work and $O(b \log n)$ depth w.h.p. for a sequence of
length $n$
Given an unsorted sequence of \elements{}, we first sort the
sequence using a comparison sort which costs $O(n \log n)$ work and
$O(\log n)$ depth~\cite{JaJa92}. Duplicate values in $S$ can now be
combined by applying a scan with $f_V$, propagating the
sum with respect to $f_V$ rightward, and keeping only the rightmost value in
the resulting sequence using a filter.

Next, we hash each \element{} to compute the set of heads and their
indices, which can be done using a parallel map and filter in
$O(n)$ work and $O(\log n)$ depth. Constructing the tails for each
head can be done in $O(n)$ work and $O(b \log n)$ depth w.h.p. by
mapping over all heads in parallel and sequentially scanning for the
tail, and applying Lemma~\ref{lem:structure}. The \plus{} is generated
similarly.  Finally, we build a purely-functional tree over the
sequence of head and tail pairs, with the heads as the keys, and the
tails as the values, which takes $O(n)$ work and $O(\log n)$ depth.

\myparagraph{Searching}
Searching (\textsc{Find}$(T, e)$) for a given
\element{} $e$ can be implemented in $O(b \log n)$ work and depth w.h.p and $O(b
+ \log n)$ work and depth in expectation. The
idea is to simply search the keys in the \treeplus{} for the first head $\leq
e$. If the head $e'$ that we find is equal to $e$ we return \textsc{true},
otherwise we check whether $e$ lies in the \tail{} associated with $e'$
sequentially and return \textsc{true} if and only if $e$ is in the tail. The
depth of the tree is $O(\log n)$ and the size of the tail is $O(b \log n)$
w.h.p. ($O(b)$ in expectation) by Lemma~\ref{lem:structure}, giving the bounds.

\myparagraph{Mapping}
Mapping (\textsc{Map}$(T, f)$) over a \treeplus{} containing $n$
\elements{} with a constant-work function $f$ can be done in $O(n)$ work and $O(b \log
n)$ depth w.h.p. We simply apply a parallel map over the underlying
purely-functional tree, which runs in $O(n)$ work and $O(\log n)$
depth~\cite{sun2018pam}. The map operation for each node in the tree
simply calls $f$ on the key (a head), and then sequentially processes
the \tail{}, applying $f$ to each \element{} in it. We then apply $f$ to
each \element{} in the \plus{}. The work is $O(n)$ as each \element{} is
processed once. As each \chunk{} has size $O(b \log n)$ w.h.p. by
Lemma~\ref{lem:structure}, the overall depth is $O(b\log n)$ w.h.p.

\myparagraph{Union Implementation}
Algorithm~\ref{alg:union-rec} first checks whether \textsc{Union} is
applicable by checking that both trees are present, and calls
\textsc{UnionBC}, which computes the union of a \plus{} and a
\treeplus{}, if either tree is $\mathsf{null}$
(Lines~\ref{line:bc1}--\ref{line:bc2}).  Next, the
algorithms calls \textsc{Expose} on $T_{2}$ to bind $k_2$ and $v_2$ (the head and its
tail) as the root of $T_{2}$, and $L_2$ and $R_2$ as $T_2$'s left and
right subtrees, respectively (Line~\ref{line:expose}). We then split
$C_{1}$ based on $k_2$ (Line~\ref{line:split}), which returns two
\treepluses{}, $B_{1}$ and $(BT_{2}, BP_{2})$ which contain all \elements{} less
than $k_2$ and all \elements{} greater than $k_2$, respectively.

Notice that some \elements{} in $v_2$, the tail from the root of $T_{2}$,
may need to be sent to the recursive call involving $B_{2}$ as
$R_{2}$'s \plus{} if a head in $B_{2}$ has a value that comes before
\elements{} in $v_{2}$. To capture these elements that should join
heads in $B_2$, we split $v_2$ based on the smallest
\element{} in $B_{2}$'s tree (Line~\ref{line:splitv2}), binding $v_L$
and $v_R$ to the lists containing \elements{} less than and greater
than the smallest \element{}, respectively.  Similarly, some of the
\elements{} in $BP_{2}$ (non-head \elements{} that are
less than all \elements{} in $BT_{2}$) may also be less
than the heads in $R_{2}$, and should therefore be merged with $k_2$'s
new \tail{}. We similarly split $BP_2$ based on the smallest element
in $R_2$ to compute $P_L$ and $P_R$ (Line~\ref{line:splitb2prefix}).
The new tail for the root is computed on Line~\ref{line:unionv2list}.
Finally, on Line~\ref{line:unionrec} we recursively call union in
parallel to obtain the \treepluses{} $C_{L}$ and $C_{R}$
respectively. Observe that $C_R$'s \plus{} must be empty because we
split the \pluses{} of the two \treepluses{} participating in the
right call to only contain \elements{} larger than the smallest head
in $B_2$ and the smallest head in $R_2$.  Therefore, all \elements{}
in both prefixes of the right recursive call will ultimately end up
joining some head's tail, implying that the prefix of $C_R$ is empty.
The tree in the output \treeplus{} is obtained by calling the
$\textsc{Join}$ function for purely functional trees on $k_2$, $v_2'$,
$C_{L}.\textsc{Tree}$, and $C_{R}.\textsc{Tree}$, and the \plus{} is
just the \plus{} from $C_{L}$ (Line~\ref{line:unionret}).


\begin{algorithm}[!t]
\caption{\textsc{UnionBC}} \label{alg:union-bc}
\small
\begin{algorithmic}[1]
\Function{UnionBC}{$C_{1}, C_2$}
\State \hspace{-1em} {\bf case} $(C_1, C_2)$\ {\bf of}
\State $((\mathsf{null}, \mathsf{null}), \_)  \rightarrow C_2$ \label{line:retc2}
\State \hspace{-0.5em}$|\ ((\_, P_1), (T_2, P_2)) \rightarrow$
\State \hspace{-0.8em}{\bf let}
\State $\mathsf{val}\ (P_{L},P_{R}) = \textsc{SplitChunk}(P_1, \textsc{Smallest}(T_2))$\label{line:bcsplitchunk}
\State $\mathsf{val}\ \codevar{keys} = \mathsf{map}(\lambda e.(\textsc{FindHead}(T_2, e), e), P_R)$\label{line:bckeys}
\State $\mathsf{val}\ \codevar{ranges} = \textsc{UniqueKeyRanges}(\codevar{keys})$\label{line:bcrange}
\State $\mathsf{val}\ \codevar{updates} = \mathsf{map}(\lambda (k, s, e).\textsc{UnionRange}(T_2, k, s, e), \codevar{ranges})$\label{line:bcupdates}
\State $\mathsf{val}\ T'_{2} = \textsc{MultiInsert}(\codevar{updates}, T_2)$\label{line:bcinsert}
\State \hspace{-0.8em}{\bf in}
\State $\mathsf{ctree}(T'_{2}, \textsc{UnionLists}(P_{L}, P_2))$\label{line:bcret}
\State \hspace{-0.8em}{\bf end}
\EndFunction
\end{algorithmic}
\end{algorithm}

\myparagraph{UnionBC}
Algorithm~\ref{alg:union-bc} implements \textsc{UnionBC}, the base-case of
\textsc{Union}, which computes the union of a \plus{} and a \treeplus{}. If
$P_{1}$ is $\mathsf{null}$, we return $C_{2}$
(Line~\ref{line:retc2}). Otherwise, $P_1$ is non-empty, and some of its \elements{}
may need to be unioned with $P_{2}$, while others may belong
in tails in $T_{2}$. We split $P_{1}$ by the first key in $T_{2}$
(Line~\ref{line:bcsplitchunk}), returning the keys in $P_1$ less than ($P_L$) and greater than ($P_R$) the first key in $T_2$.
We first deal with $P_R$, which contains \elements{} that
should be sent to $T_{2}$.
First, we find the head for each \element{} in $P_R$ in parallel by
applying a map over the elements $e \in P_R$ (Line~\ref{line:bckeys}).
Next, we compute the unique ranges for each key by calling
\textsc{UniqueKeyRanges}, which packs out the keys into a sequence of
key, start index, and end index triples containing the index of the
first and last \element{} that
found the key. This step can be implemented by a map followed by a
scan operation to propagate the indices of boundary elements, and a pack (Line~\ref{line:bcrange}).
Next, in parallel for each unique key, we call \textsc{UnionRange},
which unions the
\elements{} sent to $k$ with its current tail in $T_2$ and constructs
$\codevar{updates}$, a sequence of head-\tail{} pairs that are to be
updated in $T_{2}$ (Line~\ref{line:bcupdates}).
Finally, we call \textsc{MultiInsert} with $T_{2}$ and $\codevar{updates}$, which
returns the tree that we will output (Line~\ref{line:bcinsert}). Note that
the \textsc{MultiInsert} call here operates on the underlying purely-functional
tree. We return a \treeplus{} containing this tree, and the union of
$P_{L}$ and $P_{2}$ (Line~\ref{line:bcret}). Using the fact that the expected size
of $P_1$ is $b$, the overall work of \textsc{UnionBC} is $O(b \log
|C_2| + b\cdot b) = O(b^2 + b\log |C_2|)$ in expectation to perform
the finds and merge the \elements{} in $P_1$ with a corresponding tail.
The depth is $O(\log b \log |C_2|)$ due to the \textsc{MultiInsert}.


\begin{algorithm}[!t]
\caption{\textsc{Split}} \label{alg:split}
\small
\begin{algorithmic}[1]

\newcommand{\IndState}[1]{\State \hspace{#1em}}

\Function{Split}{$C, k$}
\State \hspace{-1em} {\bf case} $C$\ {\bf of}
\State $(\mathsf{null}, \mathsf{null}) \rightarrow (\mathsf{empty}, \mathsf{false}, \mathsf{empty})$\label{line:splitbc}
\State \hspace{-0.5em}$|\ (T, \mathsf{null}) \rightarrow$\label{line:splitplusnull}
\IndState{0} {\bf let}
\IndState{1} $\mathsf{val}\ (L, h, v, R) = \textsc{Expose}(T)$\label{line:splitexpose}
\IndState{0} {\bf in}
\IndState{1} {\bf case} $\textsc{compare}(k, h)$\ {\bf of}\label{line:splitcompare}
\IndState{2} $\mathsf{EQ} \rightarrow (\mathsf{ctree}(L, \mathsf{null}), \mathsf{ctree}(R, v))$\label{line:spliteq}

\IndState{1.5} $|\ \mathsf{LT} \rightarrow$\label{line:splitlt}
\IndState{3} {\bf let}
\IndState{4} $\mathsf{val}\ (L_{L}, (LT_{R}, LP_{R})) = \textsc{Split}((L, \mathsf{null}), k)$\label{line:splitlttree}
\IndState{3} {\bf in}
\IndState{4} $(L_{L}, \mathsf{ctree}(\textsc{Join}(LT_{R}, R, h, v), LP_{R}))$\label{line:splitltret}
\IndState{3} {\bf end}

\IndState{1.5} $|\ \mathsf{GT} \rightarrow$\label{line:splitgt}
\IndState{3} {\bf if} $(k \leq \textsc{Largest}(v))$ {\bf then}\label{line:splitgtlargest}
\IndState{4} {\bf let}
\IndState{5} $\mathsf{val}\ (v_{L}, v_{R}) = \textsc{SplitList}(v, k)$\label{line:splitv}
\IndState{4} {\bf in}
\IndState{5} $(\mathsf{ctree}(\textsc{Join}(L, \mathsf{null}, h, v_{L}), \mathsf{ctree}(R, v_R)))$\label{line:splitvret}
\IndState{4} {\bf end}

\IndState{3} {\bf else}

\IndState{4} {\bf let}
\IndState{5} $\mathsf{val}\ ((RT_{L}, RP_{L}), R_{R}) = \textsc{Split}((R, \mathsf{null}), k)$\label{line:splitrighttree}
\IndState{4} {\bf in}
\IndState{5} $(\mathsf{ctree}(\textsc{Join}(L, RT_{L}, h, v), RP_{L})$\label{line:splitrightret}
\IndState{4} {\bf end}

\IndState{3} {\bf end}

\State \hspace{-0.5em}$|\ (T, P) \rightarrow$\label{line:splithasplus}
\IndState{1} {\bf let}
\IndState{2} $\mathsf{val}\ (e_l, e_r) = (\textsc{Smallest}(P), \textsc{Largest}(P))$\label{line:splitpprefix}
\IndState{1} {\bf in}

\IndState{2} {\bf if} $k \leq e_r$ {\bf then}
\IndState{3} {\bf let}
\IndState{4} $\mathsf{val}\ (P_L, P_R) = \textsc{SplitChunk}(P, k)$
\IndState{3} {\bf in}
\IndState{4} $(\mathsf{ctree}(\mathsf{null}, P_L), (T, P_R))$
\IndState{3} {\bf end}

\IndState{2} {\bf else}
\IndState{3} {\bf let}
\IndState{4} $\mathsf{val}\ ((T_L, \_), C_R) = \textsc{Split}(T, \mathsf{null})$
\IndState{3} {\bf in}
\IndState{4} $(\mathsf{ctree}(T_L, P), C_R)$
\IndState{3} {\bf end}

\IndState{2} {\bf end}

\IndState{1} {\bf end}

\EndFunction
\end{algorithmic}
\end{algorithm}

\myparagraph{Split Implementation}
The \textsc{Split} algorithm (Algorithm~\ref{alg:split}) takes a \treeplus{}
($C$) and
a split element ($k$), and returns a pair of \treepluses{} where the first contains all
elements less than the split element, and the second contains all elements
larger than it.
It first checks to see if $C$ is empty, and returns two empty
\treepluses{} if so (Line~\ref{line:splitbc}).

Otherwise, if $C$ has a tree but not a \plus{}
(Line~\ref{line:splitplusnull}), the algorithm proceeds into the
recursive case which splits a tree. It first exposes $T$
(Line~\ref{line:splitexpose}), binding $h$ to the head at the root of
the tree, $v$ to the head's \tail{}, and $L$ and $R$ to its left and
right subtrees, respectively.
The algorithm then compares $k$ to the head, $h$. There are three
cases. If $k$ is equal to $h$ (the $\mathsf{EQ}$ case on Line~\ref{line:spliteq}), the algorithm returns a \treeplus{}
constructed from $L$ and a $\mathsf{null}$ \plus{} as the left
\treeplus{}, and $(R, v)$ as the right \treeplus{}, since all elements
in $v$ are strictly greater than $h$. Otherwise, if $k$  is less than $h$
(the $\mathsf{LT}$ case on Line~\ref{line:splitlt}), the algorithm recursively splits the \treeplus{} formed
by the left tree with a $\mathsf{null}$ \plus{}, binding $L_L$ as the left
\treeplus{} from the recursive call, and $(LT_R, LT_P)$ as the right tree and
\plus{} from the recursive call. It returns $L_L$ as the left \treeplus{}.
The right \treeplus{} is formed by joining $LT_{R}$ with the right subtree ($R$), with
$h$ and $v$ as the head and \plus{}, and taking the \plus{} as $LT_P$.
The last case, when $k$ is greater than $h$ (the $\mathsf{GT}$ case on Line~\ref{line:splitgt}) is more
complicated since $k$ can split $v$, $h$'s \tail{}. The algorithm
checks if $k$ splits $v$ (the case $k \leq \textsc{Largest}(v)$ on Line~\ref{line:splitgtlargest}), and if so
calls $\textsc{SplitList}$ on $v$ based on $k$ (Line~\ref{line:splitv}) to produce $v_L$ and
$v_R$. The algorithm returns a
\treeplus{} constructed from $L$ joined with $h$, and $v_L$ as $h$'s
tail as the left \treeplus{}, and a \treeplus{} containing $R$ and
$v_R$ as the \plus{} as the right \treeplus{}. Finally, if $k >
\textsc{Largest}(v)$, the algorithm recursively splits $R$, which is
handled similarly to the case where it splits $L$.

The last case is if $C$ has a non-$\mathsf{null}$ \plus{}, $P$. In this case, the
algorithm tries to split the \plus{}, and recurses on the tree if the
\plus{} was unsuccessfully split. The algorithm first binds $e_l$ and
$e_r$ to the smallest and largest elements in $P$. It then checks
whether $k \leq e_r$. If so, then it splits $P$ based on $k$ to
produce $P_L$ and $P_R$, which contain elements less than and greater
than $k$, respectively. It then returns a \treeplus{} containing
an empty tree and $P_L$ as the left \treeplus{}, and $T$ and $P_R$ as the right
\treeplus{}. Otherwise, $P$ is not split, but the tree, $T$ may be, and so
the algorithm recursively splits $T$ by supplying the \treeplus{} $(T,
\mathsf{null})$ to $\textsc{Split}$. Since $T$ has an empty \plus{},
splitting $T$ cannot output a left \treeplus{} with a non-empty
\plus{}. We return the recursive result, with $P$ included as the left
\treeplus{}'s \plus{}.

\myparagraph{Work and Depth Bounds}

\begin{theorem}
$\textsc{Split}(T, k)$  performs $O(b \log n)$ work and depth w.h.p.
for a \treeplus{} $T$ with $n$ \elements{}.
The result holds for all balancing schemes described
in~\cite{blelloch16justjoin}.
\end{theorem}\label{thm:split}
\begin{proof}[Proof Sketch]

As $\textsc{Split}$ is a sequential algorithm, the depth is equal to
the work. We observe that the \textsc{Split} algorithm performs $O(1)$
work at each internal node except in a case where the recursion stops
due to the split \element{}, $k$, lying between $\textsc{Leftmost}(P)$
and $\textsc{Rightmost}(P)$ (line 6), or before
$\textsc{Rightmost}(v)$ (line 15). Naively checking whether $k$ lies
before $\textsc{Rightmost}(v)$ for each tail, $v$, on a root-to-leaf
path could make us perform $\omega(b \log n)$ work, but recall that
we can store $\textsc{Rightmost}(P)$ at the start of $P$ to make
the check run in $O(1)$ work. Therefore, the algorithm performs $O(1)$
work for each internal node.

If the \treeplus{} is represented using a weight-balanced tree, AVL tree,
red-black tree, or treap then its height will be $O(\log n)$ (w.h.p.
for a treap). In the worst-case, the algorithm must recurse until a
leaf, and split the \tail{} at the leaf, which has size $O(b \log n)$
w.h.p. by Lemma~\ref{lem:structure}. Therefore the work and depth of
\textsc{Split} is $O(b \log n)$ w.h.p.
The correctness proof follows by induction and case analysis.

%
\end{proof}


\begin{theorem}
  For two \treepluses{} $T_{1}$ and $T_{2}$, the $\textsc{Union}$
  algorithm  runs in $O(b^2(k \log ((n/k) + 1)))$
  work in expectation and $O(b \log k \log n)$ depth w.h.p. where $k =
  \min(|T_{1}|, |T_{2}|)$ and $n = \max(|T_{1}|, |T_{2}|)$.
\end{theorem}\label{thm:union-app}
\begin{proof}[Proof sketch]

The extra work performed in our algorithm is due to
splitting and unioning \tails{} at each recursive call, and the work
performed in \textsc{UnionBC}. Using the fact that the expected size
of each tail is $O(b)$ we can modify the proof of the work of
\textsc{Union} given in Theorem 6 in~\cite{blelloch16justjoin} to bound our work.
In particular, we perform $O(b)$ work in expectation for each node
with non-zero splitting cost which pays at least 1 unit of cost in the
proof in~\cite{blelloch16justjoin}. To account for the work of the
\textsc{UnionBC}, observe that the dominant cost in the algorithm are
the calls to \textsc{Find} on Line 6. Also notice that calls operate on
a tree, $T$, generated by a \textsc{Split} from the parent of this
call, and the work of this step is $O(b (b + \log |T|))$ in
expectation.
We can therefore bound this work by charging each call to
\textsc{UnionBC} to the \textsc{Split} call that generated it and
applying linearity of expectations over all calls to \textsc{UnionBC}.
As we already pay $O(b \log |T|)$ for the call to \textsc{Split} in
the proof from~\cite{blelloch16justjoin} the overall work is affected
by an extra factor $b^2$, resulting in the stated work bound. Note that for
$b = O(1)$ the work is affected by a constant factor in expectation.

To bound the depth, observe that the depth of the call-tree (including
the depth of splits) can be bounded as $O(b \log n \log k)$ using the
recurrence as Theorem 8 in~\cite{blelloch16justjoin}. Furthermore, the
depth due to splitting \tails{} in recursive calls of \textsc{Union}
is $O(b \log n)$ w.h.p. per level, which is the same as the depth due
to a call to \textsc{Split}, and does not therefore increase the
depth.  Finally, although \textsc{UnionBC} can potentially have
$O((\log b +\log \log n) \log k)$ depth
due to the \textsc{MultiInsert},
\textsc{UnionBC} only appears as a leaf in the call-tree, and so its
contribution to the depth is additive. Thus the overall depth is $O(b
\log n \log k)$ w.h.p.

\end{proof}

\myparagraph{Intersection and Difference}
Lastly, for \textsc{Intersection} and \textsc{Difference}, we note
that the main difference between \textsc{Union} and
\textsc{Intersection} and \textsc{Difference} is that they may require
removing the split key (which is always maintained and joined with
using \textsc{Join} in \textsc{Union}). The only extra work is an
implementation of \textsc{Join2} over \treepluses{} which is similar
to \textsc{Join} except it does not take a key in the middle
(see~\cite{blelloch16justjoin} for details on \textsc{Join2}).

\subsection{Aspen Interface}\label{subsec:interface}

We start by defining a few types used by the interface. A
$\mathbfsf{versioned\_graph}$ is a data type that represents multiple
snapshots of an evolving graph. A $\mathbfsf{version}$ is a
purely-functional snapshot of a $\versionedgraph{}$. A $\mathbfsf{T} \
\mathbfsf{seq}$ is a sequence of values of type $\mathbfsf{T}$.
Finally, a $\mathbfsf{vertex}$ is a purely-functional vertex contained
in some $\version{}$.

\myparagraph{Building and Update Primitives}
The main functions in our interface are a method to construct the
initial graph, methods to acquire and release $\versions{}$, and
methods to modify a graph.  The remaining functions in the interface
are for traversing and analyzing $\versions{}$ and are similar to the
Ligra interface.  Aspen's functions are listed below:
\smallskip

\noindent {\textbf{\buildgraph{}}($n$ : $\mathsf{int}$, $m$ : $\mathsf{int}$,} \\
  	 \hspace*{2.30cm}$S$ : $\mathsf{int}$\ $\mathsf{seq}$\ $\mathsf{seq}$) :
     $\versionedgraph{}$\\ 
  Creates a versioned graph containing $n$ vertices and $m$
  edges. The edges incident to the $i$'th vertex are given by $S[i]$.

  \smallskip

 \noindent\textbf{\acquire{}}() : ($VG$ : $\versionedgraph{}$) : $\version{}$\\ 
  Returns a valid $\version{}$ of a $\versionedgraph{}$ $VG$.
  Note that this version will be persisted until the user calls
  \release{}.

  \smallskip

 \noindent\textbf{\release{}}() : ($VG$ : $\versionedgraph{}$, $G$ : $\version{}$) \\ 
 Releases a $\version{}$ of a $\versionedgraph{}$ $VG$.

 \smallskip

 \noindent\textbf{\insertedges{}}() : ($VG$ : $\versionedgraph{}$,
$E'$: $\mathsf{int} \times \mathsf{int}\ \mathsf{seq}$) \\ 
  Updates the latest version of the graph, $G=(V,E)$, by inserting the
  edges in $E'$ into $G$. Makes a new version of the graph equal to
  $G[E \cup E']$ visible to readers.

  \smallskip

 \noindent\textbf{\deleteedges{}}() : ($VG$ : $\versionedgraph{}$,
$E'$: $\mathsf{int} \times \mathsf{int}\ \mathsf{seq}$) \\ 
  Updates the latest version of the graph, $G=(V,E)$, by deleting the
  edges in $E'$ from $G$. Singleton vertices (those with degree $0$ in
  the new version of the graph) can be optionally removed.  Makes a
  new version of the graph equal to $G[E \setminus E']$ visible to
  readers.

  \smallskip

 \noindent\textbf{\insertvertices{}}() : ($VG$ : $\versionedgraph{}$,
$V'$: $\mathsf{int}\ \mathsf{seq}$) \\ 
  Updates the latest version of the graph, $G=(V,E)$, by inserting the
  vertices in $V'$ into $G$. Makes a new version of the graph equal to
  $G[V \cup V']$ visible to readers.

  \smallskip

 \noindent\textbf{\deletevertices{}}() : ($VG$ : $\versionedgraph{}$,
$V'$: $\mathsf{int}\ \mathsf{seq}$) \\ 
  Updates the latest version of the graph, $G=(V,E)$, by deleting the
  vertices in $V'$ from $G$. Makes a new version of the graph equal to
  $G[V \setminus V']$ visible to readers.

 \smallskip

Our framework also supports similarly-defined primitives
for updating values associated with edges (e.g., edge weights) and
updating values associated with vertices (e.g., vertex weights).
The interface is similar to
the basic primitives for updating edges and vertices.

\myparagraph{Access Primitives}
The functions for accessing a graph are defined similarly to Ligra.
For completeness, we list them below. We also provide
primitives over the $\vertex{}$ object, such as \vtxdegree{},
\vtxmap{}, and \vtxintersection{}.

\smallskip

 \noindent\textbf{\numvertices{} (\numedges{})}() : ($G$ : $\version{}$)
 : $\mathsf{int}$\\ 
 Returns the number of vertices (edges) in the graph.

 \smallskip

 \noindent\textbf{\findvertex{}}() : ($G$ : $\version{}$, $v$:
 $\mathsf{int}$) : $\{\vertex{} \cup \Box \}$\\ 
 Returns either the vertex corresponding to the vertex identifier $v$,
 or $\Box$ if $v$ is not present in $G$.

 \smallskip

 \noindent\textbf{\emap{}}() : ($G$ : $\version{}$, $U$ : \vset{}, $F$: $\mathsf{int} \times \mathsf{int} \rightarrow \mathsf{bool}$, $C$: $\mathsf{int} \rightarrow \mathsf{bool}$) : $\mathsf{\vset{}}$\\ 
 Given a \vset{} $U$, returns a \vset{} $U'$ containing all $v$ such that $(u, v) \in E$ for $u \in U$ and $C(v) = \mathsf{true}$ and $F(u, v) = \mathsf{true}$.

\subsection{Additional Experimental Results}\label{sec:additional-exps}



\begin{table}[!h]
\footnotesize
\centering
\tabcolsep=0.12cm
\hspace*{-0.2cm}
\begin{tabular}[t]{l|l | c|c|c }
  \toprule
  {\bf Application} &
  {\bf Graph} &  \textbf{Aspen Uncomp.} & \textbf{Aspen} & \textbf{(S)} \\
  \midrule
 \parbox[t]{2mm}{\multirow{3}{*}{BFS }} &
   {LiveJournal}  & 0.055 & 0.021 & 2.6x   \\ 
&  {com-Orkut}    & 0.042 & 0.015 & 2.8x   \\
&  {Twitter}      & 0.348 & 0.138 & 2.5x   \\
  \bottomrule
\end{tabular}
\caption{\textbf{Aspen Uncomp.} is the parallel time using Aspen
  with uncompressed trees, and \textbf{Aspen} is the parallel time
  of Aspen with \treepluses{} and difference encoding. \textbf{(S)} is
  the speedup obtained by Aspen over the uncompressed format. All
  times are measured on 72 cores using hyper-threading.}
\label{table:uncomp-vs-comp}
\end{table}

\begin{table*}[!t]
\footnotesize
\centering
\tabcolsep=0.12cm
\begin{tabular}[t]{l | c|c|c | c|c|c | c|c|c }
  \toprule
  {\bf Application} &  \multicolumn{3}{c|}{LiveJournal} & \multicolumn{3}{c|}{com-Orkut} & \multicolumn{3}{c|}{Twitter} \\
  & \textbf{L} & \textbf{A} & \textbf{$\frac{\mathbf{A}}{\mathbf{L}}$} & \textbf{L} & \textbf{A} & \textbf{$\frac{\mathbf{A}}{\mathbf{L}}$} & \textbf{L} & \textbf{A} & \textbf{$\frac{\mathbf{A}}{\mathbf{L}}$} \\
  \midrule
  {BFS}                & 0.015 & 0.021 & 1.40x      & 0.012 & 0.015 & 1.25x   & 0.081 & 0.138 & 1.70x          \\
  {BC}                 & 0.052 & 0.075 & 1.44x      & 0.062 & 0.078 & 1.25x   & 0.937 & 1.18  & 1.25x          \\
  {MIS}                & 0.032 & 0.054 & 1.68x      & 0.044 & 0.069 & 1.56x   & 0.704 & 0.99  & 1.40x          \\
  \midrule
  {$2$-hop}            & 3.06e-4 & 3.45e-4 & 1.13x  & 2.12e-4 & 2.52e-4 & 1.18x   & 2.79e-3 & 7.79e-3 & 2.79x  \\
  {Local-Cluster}      & 0.031   & 0.058   & 1.87x  & 0.046   & 0.097   & 2.10x   & 0.037   & 0.094   & 2.54x  \\
  \bottomrule
\end{tabular}
\caption{\small Running times (in seconds) of our algorithms over
  small symmetric graph inputs on a 72-core machine (with hyper-threading)
  where \textbf{L} is the parallel time using Ligra+, \textbf{A} is the
  parallel time using Aspen, and
  $\frac{\mathbf{A}}{\mathbf{L}}$ is the slowdown incurred by Aspen. All times are measured using 72 cores using
  hyper-threading.
}
\label{table:aspen-vs-ligra-small}
\end{table*}

\begin{table*}[!t]
\footnotesize
\centering
\tabcolsep=0.12cm
\begin{tabular}[t]{l | c|c|c | c|c|c | c|c|c }
  \toprule
  {\bf Application} &  \multicolumn{3}{c|}{ClueWeb} & \multicolumn{3}{c}{Hyperlink2014} & \multicolumn{3}{c}{Hyperlink2012} \\
  & textbf{L} & \textbf{A} & \textbf{$\frac{\mathbf{A}}{\mathbf{L}}$} & \textbf{L} & \textbf{A} & \textbf{$\frac{\mathbf{A}}{\mathbf{L}}$} & \textbf{L} & \textbf{A} & \textbf{$\frac{\mathbf{A}}{\mathbf{L}}$} \\
  \midrule
  {BFS}                & 1.68 & 3.69 & 2.19x   & 3.44 & 6.19 & 1.79x   & 8.48 & 14.1 & 1.66x \\
  {BC}                 & 14.7 & 21.8 & 1.48x   & 17.8 & 24.5 & 1.37x   & 37.1 & 58.1 & 1.56x \\
  {MIS}                & 8.14 & 12.1 & 1.48x   & 14.2 & 22.2 & 1.56x   & 32.2 & 50.8 & 1.57x \\
  \midrule
  {$2$-hop}            & 0.024 & 0.028 & 1.16x   & 0.036 & 0.038 & 1.05x   & 0.072 & 0.075 & 1.04x \\
  {Local-Cluster}      & 0.013 & 0.020 & 1.53x   & 0.013 & 0.021 & 1.61x   & 0.016 & 0.024 & 1.50x \\
  \bottomrule
\end{tabular}
\caption{\small Running times (in seconds) of our algorithms over
  large symmetric graph inputs on a 72-core machine (with hyper-threading)
  where \textbf{L} is the parallel time using Ligra+, \textbf{A} is the
  parallel time using Aspen, and
  $\frac{\mathbf{A}}{\mathbf{L}}$ is the slowdown incurred by Aspen. All times are measured using 72 cores using
  hyper-threading.
}
\label{table:aspen-vs-ligra-large}
\end{table*}


\newcommand{\STAB}[1]{\begin{tabular}{@{}c@{}}#1\end{tabular}}

\end{document}